\documentclass[pre,twocolumn,floatfix]{revtex4-2}

\usepackage{amsfonts, amssymb, amsmath,amsthm}
\usepackage{xcolor}
\usepackage{graphicx}
\usepackage{hyperref}
\usepackage{tabularx}
\usepackage{subfigure}
\usepackage[normalem]{ulem}

\begin{document}
\title{Forecasting landslides using community detection on geophysical satellite data}

\author{Vrinda D.~Desai}
\author{Farnaz Fazelpour}
\affiliation{Physics Department, North Carolina State University, Raleigh, NC, USA}
\author{Alexander L. Handwerger}
\affiliation{Joint Institute for Regional Earth System Science and Engineering, University of California Los Angeles, Los Angeles, CA, USA and Jet Propulsion Laboratory, California Institute of Technology, Pasadena, CA, USA}
\author{Karen E.~Daniels}
\affiliation{Physics Department, North Carolina State University, Raleigh, NC, USA~}

\date{\today} % Leave empty to omit a date

\begin{abstract}
As a result of extreme weather conditions, such as heavy precipitation, natural hillslopes can fail dramatically; these slope failures can occur on a dry day due to time lags between rainfall and pore-water pressure change at depth, or even after days to years of slow-motion. While the pre-failure deformation is sometimes apparent in retrospect, it remains challenging to predict the sudden transition from gradual deformation (creep) to runaway failure. We use a network science method -- multilayer modularity optimization -- to investigate the spatiotemporal patterns of  deformation in a region near the 2017 Mud Creek, California landslide. We transform satellite radar data from the study site into a spatially-embedded network in which the nodes are patches of ground and the edges connect the nearest neighbors, with a series of layers representing consecutive transits of the satellite.  Each edge is weighted by the product of the local slope (susceptibility to failure) measured from a digital elevation model and ground surface deformation (current rheological state) from interferometric synthetic aperture radar (InSAR). We use multilayer modularity optimization to identify strongly-connected clusters of nodes (communities) and are able to identify both the location of Mud Creek and nearby creeping landslides which have not yet failed. We develop a metric, community persistence, to quantify patterns of ground deformation leading up to failure, and find that this metric increases from a baseline value in the weeks leading up to Mud Creek's failure. These methods  promise as a technique for highlighting regions at risk of catastrophic failure.
\end{abstract}


\maketitle

\section{Introduction}

Slow-moving landslides creep downhill at rates from millimeters to several meters a year \citep{lacroix_life_2020}, a process that sometimes proceeds for hundreds of years \citep{hu_four-dimensional_2020, rutter_quantifying_2011}. Yet, a slow-moving landslide can reach runaway acceleration, and in many cases this is likely to be triggered by heavy precipitation, as happened in the 2017 Mud Creek landslide located on the Big Sur Coast, CA.   The processes which influence these transitions are myriad, and additionally include topographic slope, hydrological and mechanical properties, and natural or man-made disturbances. Advancement in remote sensing techniques allows for the detection of ground deformation at spatiotemporal resolutions that were not achievable a decade ago \citep{aryal_displacement_2012, tordesillas_spatiotemporal_2021}. Previous work in landslide forecasting has focused on either (1) estimating the location and shape of a landslide using spatial analysis of slope stability \citep{carla_perspectives_2019} or (2) the time of failure using temporal analysis of ground deformation \citep{intrieri_forecasting_2019}. There has been recent progress on forecasting both location and time \citep{singh_spatiotemporal_2020, tordesillas_spatiotemporal_2021}. Linking kinematic data and the underlying micro-mechanics of granular failure together has shown to be effective in forecasting the location and time of granular failure \citep{tordesillas_spatiotemporal_2021, tordesillas_data-driven_2018, cascini_forecasting_2022}.

Interferometric processing of synthetic aperture radar (InSAR) makes studies of landslide and other ground deformation possible at high spatial and temporal resolution. Ground deformation over large surface areas of the Earth is measured with centimeter to millimeter accuracy using InSAR \citep{gabriel_mapping_1989} by exploiting the phase difference between two SAR snapshots acquired over the same area at two different times. This quantifies the amount of ground deformation during that time interval, projected along the radar line-of-sight (LOS) direction \citep{franceschetti_synthetic_2018, di_traglia_joint_2021}. InSAR techniques have been applied in the study of landslides and are starting to be incorporated into early warning systems (EWS)\citep{teshebaeva_deepseated_2019, teshebaeva_alos/palsar_2015, carla_integration_2018,carla_integration_2018, di_traglia_joint_2021, carla_perspectives_2019}. 

EWS have used these advancements in radar monitoring data to monitor areas of known risk and identify regions heavily affected by disaster \citep{bernardi_use_2021}, but there are still significant drawbacks and limitations to the forecasting of landslides and other hazards \citep{tordesillas_data-driven_2018}. One such drawback is the use of thresholds that are defined to predict the onset of failure in landslides \citep{dick_development_2015}, and recent studies have shown that these thresholds are site-specific and subjective \citep{tordesillas_data-driven_2018}. A common practice is to only consider the region that is accelerating rapidly \citep{dick_development_2015}, but important information about the interactions occurring between the region and the surrounding area then go undetected \citep{tordesillas_spatiotemporal_2021, zhou_pinpointing_2022,das_near_2019}. 

The transition of a creeping landslide to a catastrophic failure is complex and the system's behavior emerges from dynamics that are occurring beneath the surface \citep{agliardi_slow--fast_2020}. The landslide system is modeled as a collection of patches that are interacting with each other in such a way that the emerging behaviour is not immediately predictable \citep{torres_why_2021}, which is why it is crucial to not only look at the immediate region, but also the surrounding area. Thus, one of the biggest challenges to EWS in regards to landslides is harnessing the full potential of these data sets using the analysis and interpretation of the underlying micro-mechanics in the time before granular failure \citep{tordesillas_spatiotemporal_2021}. 

Recent studies have shown that granular failure does not occur spontaneously, but instead there is a transitional period between stable deformation and catastrophic failure, which is indicated by a distinct dynamical pattern, influenced by kinematics\citep{singh_spatiotemporal_2020, tordesillas_data-driven_2018, le_bouil_emergence_2014}. 
Consider a region that consists of multiple closely connected patches, or clusters. As the region approaches failure, the spatiotemporal dynamics have been characterized into two pieces: (1) similarity within a cluster increases as they begin to move together and (2) similarity between clusters decreases \citep{singh_spatiotemporal_2020}. The transition to imminent failure is guided by the spatiotemporal persistence of community $C$, followed by a sharp increase in the motion, or velocity, of $C$ in respect to the rest of the system, or area \citep{singh_spatiotemporal_2020}.

Our goal is to use kinematic data retrieved from remote sensing techniques to find patches of ground, defined as communities, that are likely to fail and their impending time of failure. To develop a holistic approach that integrates the dynamics of ground deformation, along with the physics of slope stability, we use network science techniques of micro-scale granular failure (grains on the order of 1 mm to 1 cm) \citep{fazelpour_failure_2022,mei_micro-_2022, tordesillas_force_2007} and apply it on a macroscale system (patches of area on the order of 7x7 to 20x20 m) \citep{singh_spatiotemporal_2020}. A network framework gives us the ability to infer the structure and underlying dynamics of a creeping landslide using information such as displacement that can be retrieved using remote sensing techniques \citep{torres_why_2021}. 

\subsection{Study Site: Mud Creek vicinity}

Within the California Coast Ranges, landslides are abundant due to mechanically weak rocks, active uplift, and high seasonal precipitation \citep{handwerger_shift_2019,roering_beyond_2015, handwerger_kinematics_2015, scheingross_fault-zone_2013}. More than 650 active slow-moving landslides have been identified and mapped within the region, but few of them have failed catastrophically \citep{scheingross_fault-zone_2013, handwerger_kinematics_2015}. 
On a dry day (May 20, 2017) following a long period of heavy rainfall that ended on April 17, 2017, the Mud Creek landslide, which had been slowly moving for decades, failed catastrophically and destroyed part of California State Highway 1 (CA-1) \citep{handwerger_shift_2019, warrick_characterizing_2019}.

\begin{figure}
    \centering
    \includegraphics[width=\linewidth]{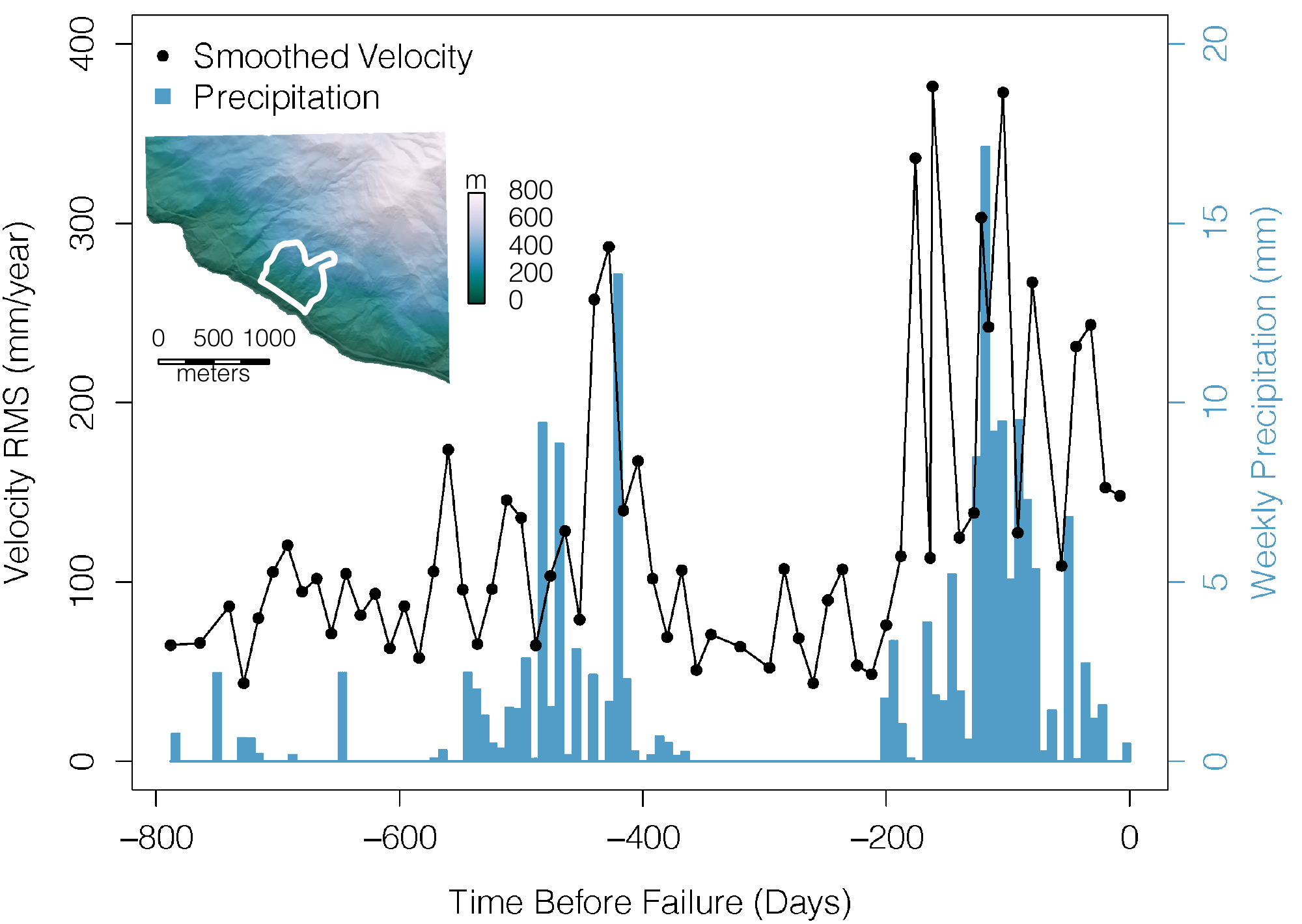}
    \caption{The root mean square  (RMS) of the line of sight (LOS) velocity (black circles) for the entire study area leading up to the day of failure ($T=0$); data from \citet{handwerger_shift_2019}. Blue bars are the total weekly precipitation, from PRISM \citep{noauthor_prism_2015}. The wet seasons correspond to roughly December to May each year.}
    \label{fig:prcpvel}
\end{figure}

The Central Coast Ranges, which contain Mud Creek, receive about $\sim 1~\mathrm{m}$ of rain per year, of which 80\% falls between October and May \citep{handwerger_shift_2019, swain_increasing_2018}. The precipitation in this region varies significantly from year to year due to changes in the frequency and strength of atmospheric rivers; climate models predict that California will experience a dramatic increase in frequency of shifts from extreme dry seasons to extreme wet seasons over the next couple of decades \citep{swain_increasing_2018}. Since landslide dynamics can be driven by seasonal changes, as seen in Fig.~\ref{fig:prcpvel}, the predicted increase in changes from dry to wet seasons is a cause for concern \citep{handwerger_shift_2019, swain_increasing_2018, allen_nino-like_2017}. 

Changes in the shear strength of a hillslope are dependent on the water-saturation level of the material \citep{iverson_acute_2000, mitarai_wet_2006, schulz_relations_2009}. If the material is not fully saturated, then capillary effects increase shear strength and help stabilize a hillslope \citep{mitarai_wet_2006, lu_hillslope_2013, iverson_acute_2000}. On the other hand, if the material is almost or fully saturated, then an increase in the pore-water pressure leads to a reduction in the shear strength of the material \citep{mitarai_wet_2006, schulz_relations_2009}, and may possibly lead to instability within the hillslope \citep{lu_hillslope_2013, schulz_relations_2009, iverson_acute_2000, agliardi_slow--fast_2020}. Note that the pore-water pressure varies strongly with depth, as rain permeates the hillslope over time. In addition, the parameters that influence slope stability originate from critical-state soil mechanics with a Mohr-Coulomb failure criterion, which assumes that failure is controlled by maximum shear strength \citep{girard_failure_2010, jaeger_fundamentals_2007}. Saturated material, corresponding to high pore-water pressure, plus steep slopes increases the susceptibility to slope failure \citep{agliardi_slow--fast_2020}. 

In this paper, we examine the area in and around Mud Creek ($3~\mathrm{km} \times 3~\mathrm{km}$), shown in Fig.~\ref{fig:prcpvel}, to explore the dynamics within Mud Creek and compare it to the rest of the area, which consists of 2 other creeping landslides that did not fail. We are interested in analyzing, and with the eventual aim of forecasting, the transition of a slow-moving landslide to sudden catastrophic failure.  

\subsection{Comparing Mud Creek to Other Regions}

\begin{figure}
    \centering
    \includegraphics[width=\linewidth]{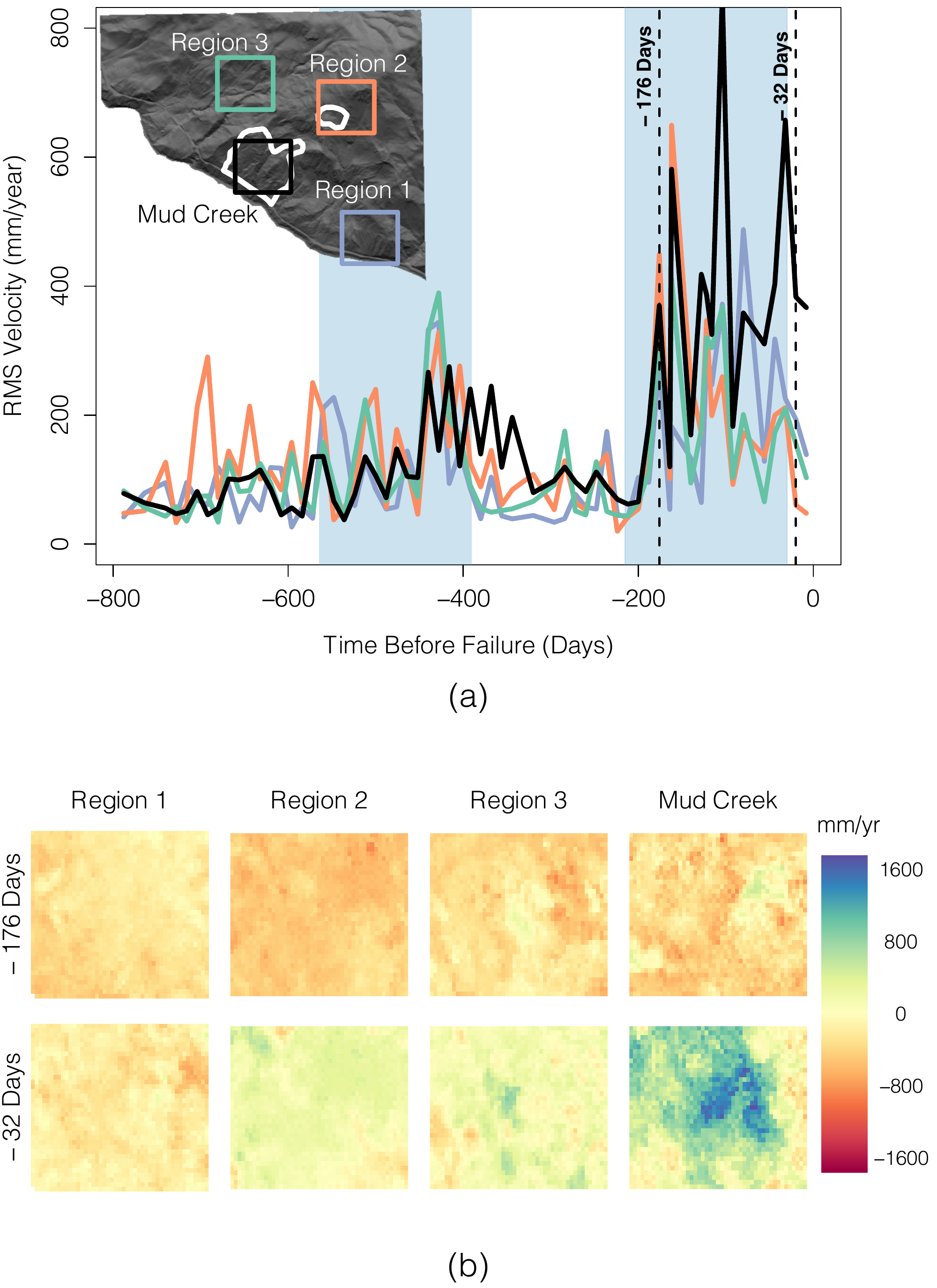}
    \caption{(a) The root mean square  (RMS) of the line of sight (LOS) velocity, for each of the  4 regions. The wet seasons shown in Fig.~\ref{fig:prcpvel} are shaded in blue. Inset: bare earth hillshade map from lidar showing the four regions within the study area. (b) Maps of the LOS velocity for each of the 4 regions at two times: $-32$ and $-176$ days before failure. Data is from \citep{handwerger_shift_2019}.}
    \label{fig:velregions}
\end{figure}

Since there is only one creeping landslide that catastrophically failed within the study area, we create a set of controls to test for sensitivity and specificity. For this, we additionally selected three regions within the study area to analyze alongside Mud Creek (see Fig.~\ref{fig:velregions}). Of these, Region 2 is a creeping landslide that was not identified in \citet{handwerger_shift_2019}, but instead was recently observed with airborne InSAR data (see Fig.~\ref{fig:velregions}); and Regions 1 and 3 are selected because they are steep areas that are not creeping. For each of these regions, the LOS velocity (from Sentinel-1 InSAR data, described in more detail in \S\ref{sec:Sentinel}), plotted in Fig.~\ref{fig:velregions}, shows that there is not a significant difference between them and Mud Creek for the majority of the time period. In Fig.~\ref{fig:velregions}(b), the velocities for the four regions are mapped at two different time steps ($-176$ and $-32$ days) to show that the regions are moving similarly to each other at $-176$ days and later at $-32$ days Mud Creek is moving significantly faster than the other three regions. 

\subsection{Networks}

It is difficult to quantify the structural changes happening beneath the surface, and thereby forecasting when and where a landslide is likely to occur. Thus, it is desirable to turn to reduced descriptions of the system. One such technique, network science, has been successful at quantifying and understanding the spatiotemporal dynamics of complex systems without the inclusion of a detailed model of the internal interactions. Previous studies, such as \citep{palla_quantifying_2007, whitaker_colloidal_2019}, have shown that a network clustering technique known as \textit{community detection} (reviewed in \citet{fortunato_community_2010}) can be used to extract hidden relations among nodes in static single-layer networks. Recent studies, such as the analysis of the evolution of the force networks within granular materials by \citet{papadopoulos_evolution_2016}, have extended that technique to be used in \textit{multilayer networks} to analyze spatially-embedded networks ({\it i.e. networks in which nodes only share edges with their own neighbors)} that are more similar to earth materials. 

\begin{figure*}
    \centering
    \includegraphics[width=\linewidth]{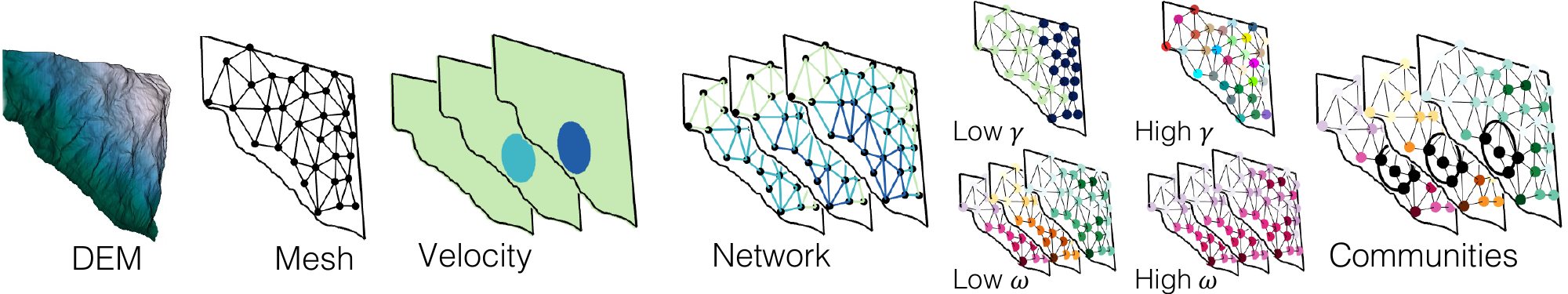}
    \caption{Schematic of the construction and analysis of our multilayer network. We sample the digital elevation model (DEM) and a series of velocity maps on a disordered mesh to create a multilayer network weighted by both. Using this network, we selected the spatial coupling parameter $\gamma$ and the temporal coupling parameter $\omega$ to detect communities of interest.}    
    \label{fig:schematic}
\end{figure*}

Network science provides a way to characterize this complex system for consistent patterns of causality via community-detection techniques. For each snapshot of ground deformation, there is a correlating static network layer that is connected together into a multilayer network, where a simplistic version of a multilayer network is shown in Fig.~\ref{fig:schematic}. We analyze the transition of a slow-moving landslide, as represented in a multilayer network, to sudden catastrophic failure using \textit{community persistence}, a measure we have developed to quantify the consistency of community structures. Using this measure, we are able to narrow down the period in which Mud Creek could be said to be at risk of failure, compared to what would be possible using InSAR alone.

\section{Data Assimilation}

We use two types of data to create our multilayer network: digital elevation models (DEM) for spatial information and slope and displacement snapshots from InSAR to estimate the current rheological state. This section details the process by which we define our multilayer network from the DEM and InSAR data; the edges within each time step (layer $\ell$) are weighted by both the local slope (steeper slopes are more susceptible to failure) and the local creep velocity (faster speeds indicate a current soil state with more fluid rheology). Local slope is held fixed over the duration of the analysis.

\subsection{Digital Elevation Model \label{sec:DEM}}

We measure the elevation at all positions in the study site using a Digital Elevation Model (DEM) at 1~m resolution, downloaded from OpenTopography (May 17 2021), chosen because it penetrates vegetation and measures the true ground surface. This DEM is from the USGS, 3D Elevation Program: ARRA-CA CentralCoast-Z3 2010 \cite{noauthor_usgs_2010}. We define the elevation as a field $h({\vec r})$ with position $\vec{r} = (x,y)$ given in UTM coordinates. From this, we calculate the slope edge weight as $s = \frac{|\Delta h|}{|\Delta \vec{r}|}$, which has a flat distribution for $0^{\circ} \leq s \leq 18^{\circ}$, after which steeper slopes are rarer, up to a maximum of $32^\circ$ (such as within Mud Creek). The mean slope weight for the other three regions are: $25^\circ$, $24^\circ$, $25^\circ$. Note that the study site contains steeper slopes, but those are averaged out when calculating the edge weights.

\subsection{InSAR \label{sec:Sentinel}}
We utilize an existing dataset from \citet{handwerger_shift_2019}, comprising InSAR snapshots from the Copernicus Sentinel-1 A/B satellites (see \citet{handwerger_shift_2019} for InSAR processing details). The resulting series of snapshots provides a total of 63 layers with $10~\mathrm{m} \times 12~\mathrm{m}$ resolution estimating ground deformation in a $3~\mathrm{km} \times 3~\mathrm{km}$ region in the vicinity of Mud Creek area from February 2015 to May 2017. Each InSAR snapshot provides LOS displacements, where the direction of motion is either towards or away from the satellite. Using the InSAR data, we have a layer $\ell$ for each snapshot within the multilayer network. These layers span 6, 12, or 24 days due to either missed orbits (repeat cycle is 6 days)  or low coherence.

For each layer $\ell$ (corresponding to an InSAR snapshot), we defined displacement as a field $U(x,y,\ell)$. Cumulative displacement is the displacement of the surface from the first reference frame ($\ell = 0$), for which $U(x,y,0) = 0$ everywhere. Subsequent layers contain the cumulative displacement that has happened during layers $0$ to $\ell$. To quantify deformation, we calculate the incremental displacement $u$ between pairs of adjacent snapshots as $u(x,y,\ell) = U(x,y,\ell) - U(x,y,\ell-1)$. Since the acquisition interval for the satellites varied from 6, 12, and 24 days, we converted incremental displacement into velocity, $v(x,y,\ell) =\frac{u(x,y,\ell) - u(x,y,\ell-1)}{\Delta{t}}$, where $\Delta{t}$ is time between any 2 consecutive snapshots.

The noise from the InSAR data (not visualized in Fig.~\ref{fig:schematic}) increases when we take the derivative of the LOS displacement; this obscures motion that is otherwise detected when retroactively looking at cumulative displacement maps. In order to reduce noise within each layer of the original dataset, we apply a  a  $3\times 3$ cell matrix of a Gaussian smoothing kernel.

\subsection{Defining the multilayer network}

Our analysis takes place on a  multilayer network, where each layer $\ell$ represents a time in space (corresponding to an InSAR snapshot). Each layer is represented as a spatially embedded network consisting of nodes and edges. Each node represents a patch of area, and each edge is the connection between patches of areas that are neighbors. This network consists of the same nodes and edge connections for all layers within the system, which we define as a mesh.

To create the spatially embedded network, we sampled the DEM at $n \sim 5000$ nodes for the study area shown in Fig.~\ref{fig:velregions}, selected to be large enough to be computationally efficient but small enough to provide spatial resolution within ridges and valleys. We considered uniform, random, and uniform random grids to create a mesh that would not introduce artefacts into the results or lead to mesh-specific results. Using Poisson disc sampling \citep{cook_stochastic_1986}, a type of uniform random grid, we randomly selected node locations spaced an average of 20 m apart, without any preferred orientation, as shown in Fig.~\ref{fig:schematic}. Pairs of adjacent nodes are connected by edges using Delaunay triangulation (selects the closest nodes to each node), defining our binary adjacency matrix $B_{ij}$. We created an ensemble of 10 such meshes that are uniformly sampled across the area while also allowing us to test for sensitivity of the results to the mesh that was used. 

Information about the current state of rheology and the susceptibility of failure was incorporated into the network using data from the DEM and InSAR to weight the edges. For each edge ${ij}$, we calculated the slope $s_{ij} = \frac{|h_i - h_j |}{|\vec{r}_i - \vec{r}_j|}$, where $h_i$ is the average elevation (given by the DEM) of the patch of area represented by node $i$ and $\vec{r}_i$ is the central coordinates of the patch of area, or node $i$. For an InSAR snapshot (layer $\ell$), each node $i$ has some velocity $v_{i\ell}$ that represents the backward-difference velocity, selected since it is convenient for use in real-time monitoring. We then calculated velocity for each edge ${ij}$ as the absolute average velocity defined as $|{\bar v}_{ij\ell}| = |\frac{v_{i\ell} + v_{j\ell}}{2}|$. The edges are weighted as $s_{ij} |{\bar v}_{ij\ell}|$ for each layer $\ell$, where $s_{ij}$ is constant for all layers, but $|{\bar v}_{ij\ell}|$ changes for each layer $\ell$. We define our adjacency matrix as
 \begin{equation}
    A_{ij\ell} = s_{ij} |{\bar v}_{ij\ell}| B_{ij}
      \label{eqn:adjmat}
 \end{equation}
for a single layer $\ell$ as a snapshot in time, with node indices ${i,j}$ specifying spatial location.

Our computer code is available on GitHub \cite{github_desai}.

\section{Network Analysis}

A \textit{community} within a network is a collection of nodes that are more strongly related to each other than to other nodes (see Fig.~\ref{fig:schematic}) \citep{porter_communities_2009, fortunato_community_2010, mucha_community_2010}. We seek to identify whether communities are present within the Mud Creek dataset, and evaluate whether there are particularly unstable regions or periods within the system. For example, clusters of nodes that are connected by edges, and moving at relatively higher speeds and/or on steeper slopes, could be identified as a community. 
The community structure of a multilayer network shows the evolution of the communities over time, where communities are formed from partitioning the graph based on a defining characteristic \citep{porter_communities_2009}.

\subsection{Community detection}

In this study, we analyze our multilayer network (Eq.~\ref{eqn:adjmat}) using a method based on optimizing the modularity $Q$; the method partitions nodes into communities (each community is thereby given an arbitrary ID by the algorithm to identify it) where the edge weights within the community are stronger than one would expect in comparison to a null model \citep{porter_communities_2009, newman_finding_2004,newman_modularity_2006}. We optimize the modularity of the community structures using the NetWiki GenLouvain implementation for multilayer networks \citep{netwiki,mucha_community_2010}. The multilayer modularity is calculated as
  \begin{equation}
     Q
     = \frac{1}{2\mu}\sum_{ij\ell m} [(A_{ij\ell} - \gamma P_{ij\ell})\delta_{\ell m} + \omega_{j\ell m}\delta_{ij}]\delta(c_{i\ell}, c_{jm}),
     \label{eqn:modularity_multi}
 \end{equation} 
where $\mu$ is the sum of all edge weights in the network, $c_{i\ell}$ is the community of node $i$, $c_{j\ell}$ is the community of node $j$, $\gamma$ is the \textit{structural resolution parameter}, $\omega_{j\ell m}$ is the \textit{interlayer coupling}, and $P_{ij\ell}$ is the expected edge weight between nodes $i$ and $j$ under a specified \textit{null model} in layer $\ell$ \citep{mucha_community_2010}. Note that using a global null model disregards the temporal fluctuations and does not improve the results of the community detection.

To optimize the modularity, the nodes are partitioned into communities in such a way that the total connection strength --- the cumulative weight of connected edges --- within a community is much greater than what would be expected from a null model. For any network, the choice of null model provides the expected weight of a randomly-selected pair of nodes, and affects the interpretation of the network \citep{sarzynska_null_2016}. Because our system is a spatially-embedded network, the work of Bassett {\it et al.} \cite{bassett_extraction_2015, papadopoulos_evolution_2016} guides our choice of the \textit{geographical null model} rather than the more common Newman-Girvan null model \citep{newman_finding_2004,newman_modularity_2006} for which all edges are equally connected to each other. The geographical null model 
is defined as 
\begin{equation}
     P_{ij\ell} = \rho_\ell B_{ij},
     \label{eqn:geographic null model}
\end{equation}
where $\rho_\ell = \langle s_{ij} |v_{ij\ell}| \rangle_{ij}$ is the mean edge weight of network layer $\ell$ (averaged over all ${ij}$ edges) and $B_{ij}$ is the binary adjacency matrix, which is constant through all layers and encodes the geophysical proximity. 

There are two resolution parameters used to calculate modularity $Q$, both of which are selected to control the total number of communities detected, as seen in Fig.~\ref{fig:schematic}. The structural resolution parameter $\gamma$ sets the weight of the geographic null model. Within each layer $\ell$, $\gamma$ determines the spatial resolution at which nodes are partitioned into communities, via comparison to the mean edge weight $\rho_\ell$;  higher values of $\gamma$ will yield more communities that are smaller in size \citep{papadopoulos_evolution_2016}.
The corresponding temporal resolution parameter is contained within the interlayer coupling $\omega_{j\ell m}$, which is the weight of the edge coupling node $j$ to itself in layers $(\ell, m)$. Here, we take $\omega_{j\ell m} = \omega$ (constant), for simplicity. We want to consider $\omega$ values that take into account the strain-loading history. Higher values of $\omega$ more strongly favor the continuity of communities across the series of layers, while for $\omega = 0$, each layer is considered independently (no communities span the layers), and thus no consideration of the strain-loading history.

\subsection{Selecting the coupling parameters ($\gamma$, $\omega$) \label{sec:parameters}}

We wish to select resolution parameters that highlight communities within the network which (1) have stronger weights than average and (2) continue that behavior over some time interval. 
To satisfy the former, we pick $\gamma =1$; this value favors communities with stronger-than-average weights (larger than $\rho_\ell$) \citep{papadopoulos_evolution_2016}. In contrast, the interlayer coupling $\omega$ is normally of order one \citep{papadopoulos_evolution_2016}. 
For the latter, we want to find a value of $\omega$ which is large enough that important communities persist over time, but not so large that we fail to detect changes in the temporal state of the study area. 
Since there is not yet consensus in the literature on
specific values of $\omega$, we explore two measures which have been useful in earlier studies:\textit{flexibility} and \textit{stationarity} \citep{papadopoulos_evolution_2016, bassett_dynamic_2011, palla_quantifying_2007}. 

We quantify the local properties of the nodes using a measure known as flexibility \citep{bassett_dynamic_2011}. It is the measure of the number of layers, $g_i$, that each node $i$ changes community allegiance over the total possible layers ($L-1$). The flexibility of the multilayer network as a whole for all nodes $N$ is defined as:
\begin{equation}
    \Xi = \frac{1}{N}\sum_i\frac{g_i}{L-1}.
    \label{eqn:flexibility}
\end{equation}

To capture the global properties of the communities found within the network, we measure the consistency with which nodes are placed in communities over time using a measure known as stationarity \citep{bassett_robust_2013, palla_quantifying_2007}. As given in \citet{papadopoulos_evolution_2016}, the stationarity of a multilayer network is $\zeta$ for all nodes $N$:
\begin{equation}
    \zeta = \frac{1}{C} \sum_c \frac{\sum_{\ell_i}^{\ell_f-1} J(c_\ell, c_{\ell+1})}{\ell_f -\ell_i - 1},
    \label{eq:stationarity}
\end{equation}
where $C$ is the total number of communities and $J(c_\ell, c_{\ell+1})$ is the joint fraction (the ratio of the number of mutual nodes in community $c$ in layers $\ell$ and $\ell+1$ to the number of nodes in the union of layers $\ell$ and $\ell+1$). The stationarity $\zeta$ approaches 1 at high values of $\omega$. In our calculation, we exclude communities that never contain more than one node \citep{papadopoulos_evolution_2016}.

\begin{figure}
    \centering
    \includegraphics[width = \linewidth]{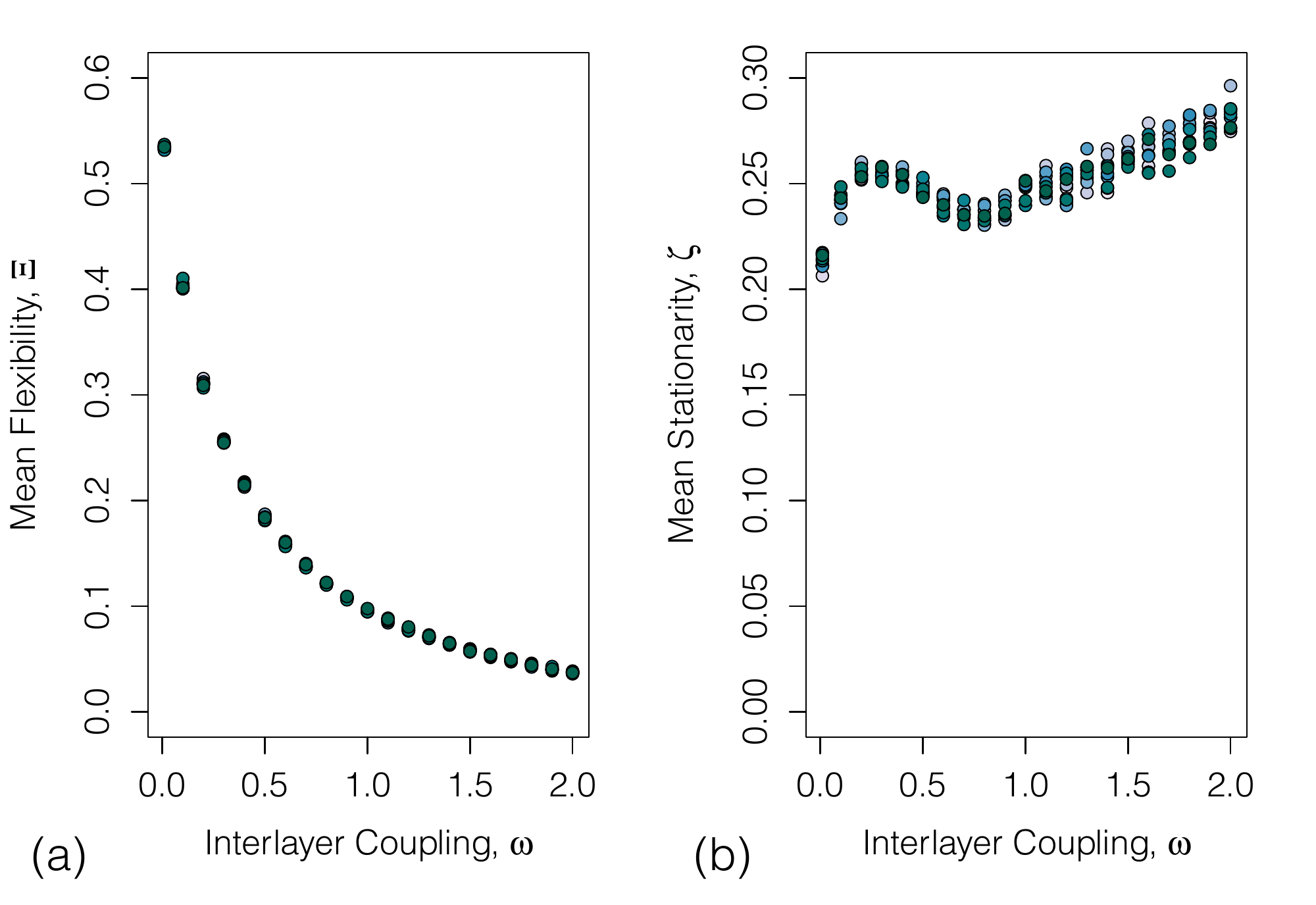}
    \caption{(a) Network flexibility $\Xi$  (Eq.~\ref{eqn:flexibility})  and (b) network  stationarity $\zeta$  (Eq.~\ref{eq:stationarity}) as a function of the choice of a particular temporal coupling parameter $\omega$, for fixed spatial coupling $\gamma = 1$. Each data point is for one of the 10 meshes, distinguished by color.}    \label{fig:stationflex}
\end{figure}

Fig.~\ref{fig:stationflex} shows the values of both the flexibility (Eq.~\ref{eqn:flexibility}) and the stationarity (Eq.~\ref{eq:stationarity}), repeated for each of the 10 meshes. In order to be sensitive to both the intra- (spatial) and inter- (temporal) layer connections, $\omega$ should also be on the same order of magnitude as $\gamma$; therefore, we explore values of $\omega \in [0,2]$. We observe that as $\omega$ increases,  $\Xi$ decreases and approaches zero;  for $\omega \in [0.1,1.6]$, the differences between the meshes are minimal, and the dependence on $\omega$ is weak ($\zeta = 0.25 \pm .01$ throughout).

Using these plots, we consider a few things when picking $\omega$. Low flexibility and/or high stationarity would result in communities that are insensitive to real temporal fluctuations which inform the forecasting; the temporal edge weights dominate for $\omega > 1$. On the other end of the spectrum, high flexibility and/or low stationarity would mean that the strain loading is not being taken into account; the spatial edge weights dominate the communities for $\omega < 1$. Therefore, we select $\omega = 1$ because it balances the strain-loading history (inter-) with the rheological properties (intra-layer).

\section{Results}

We aim to determine whether we have identified the (known) geometry and location of Mud Creek with the results of the community detection algorithm (explored below). Once this is achieved, we examine the details of the degree of community persistence over time, and use our observations to construct a measure which is able to forecast increasing risk of failure.

\subsection{Quantifying community structures}

\begin{figure}
    \centering
    \includegraphics[width = \linewidth]{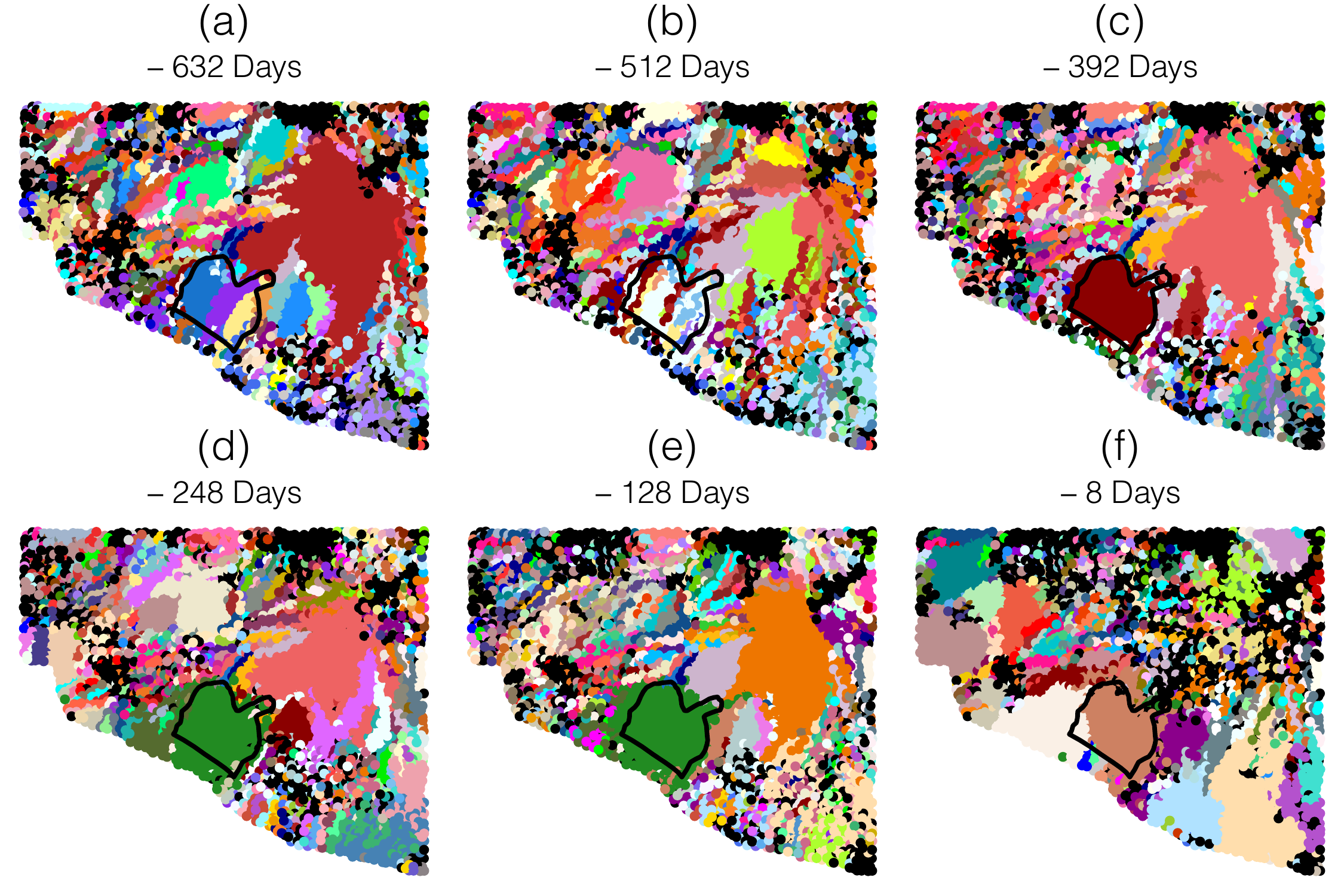}
    \caption{Results of community detection using  $\gamma$ = 1, $\omega = 1$, shown for six representative layers prior to failure at $T=0$ days. Nodes that are members of the same community have the same (arbitrary) color, corresponding to a single community ID number. Black nodes represent singleton communities. The black outline is Mud Creek.}
    \label{fig:communities}
\end{figure}

In order to identify areas of concern, we run community detection using the modularity function defined in Eq.~\ref{eqn:modularity_multi} with parameters $\gamma = 1$ and $\omega = 1$. For each InSAR snapshot, there is a resulting community structure: 62 in total, 6 of which are shown in Fig.~\ref{fig:communities}. We only present the results of one mesh in Fig.~\ref{fig:communities}, but this is representative of the results for all 10 meshes, indicating that the algorithm is not sensitive to the particular mesh used. Mud Creek is identified as a community early on, clearly showing up in Fig.~\ref{fig:communities}(c) in red and remains coherent for the next $\sim$ 400 days prior to its ultimate failure.

We first consider a simple metric: how the number of communities found within each layer changes over time. In Fig.~\ref{fig:communities}, the number of singleton communities (black nodes) increases as Mud Creek approaches failure, while the number of nonsingleton communities remains stable. Singleton communities most commonly appear at the border of communities; sometimes they join a community, and sometimes they arise at new boundaries when a community splits apart. As we will see later, these dynamics are connected to changes in persistence of community structures as Mud Creek approaches failure at $T=0$.

\begin{figure}
    \centering
    \includegraphics[width=\linewidth]{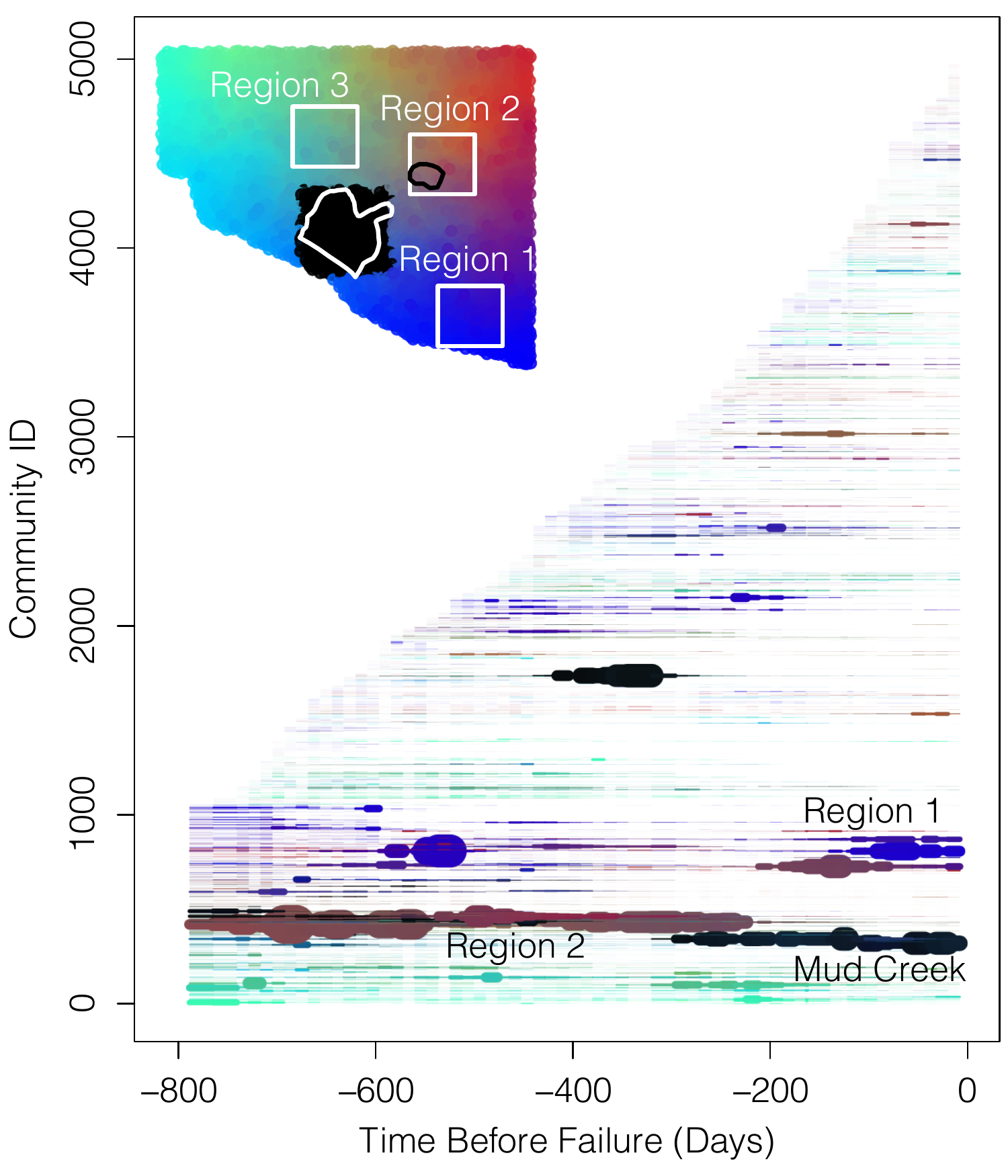}
    \caption{Strip-chart showing the evolution of the community structures detected for parameters $\gamma = 1$ and $\omega = 1$, separately highlighting location, duration, and significance $\Psi_c$. Each detected community found has an ID number that is arbitrary (but corresponds to the order of detection), and is plotted  as a line with three properties: color specifies its location (see inset map), the extent along the horizontal axis specifies its duration, and the thickness of of the line measures the size and strength of that community at layer $\ell$, according to Eq.~\ref{eq:thickness}. The outlines of the creeping landslides found within the area are shown in the inset map.}
    \label{fig:communityvTime}
\end{figure}

To identify the underlying pattern within these identified communities, we visualize the evolution of the community structures in Fig.~\ref{fig:communityvTime}(a), where each horizontal line represents a single community, whose ID is assigned in order of its identification by the algorithm, and the color of the line represents the approximate location of that community (see inset). 

There are two characteristics of a community that we considered: size and strength. The \textit{strength} of a community is the mean of the weights of the edges, and the \textit{size} of a community is the number of nodes. We find that a joint measure of community size and strength, normalized by global values, is useful for highlighting the significance  $\Psi_{c, \ell}$ of particular  communities detected, calculated as
\begin{equation}
 \Psi_{c,\ell} = \frac{n_{c, \ell}}{N} \frac{S_{c, \ell}}{S_{\ell}},
 \label{eq:thickness}
\end{equation}
where $n_{c, \ell}$ is the number of nodes in community $c$ at layer $\ell$, $N$ is the total number of nodes, and $S_{c, \ell}$ and $S_\ell$ is the strength of community $c$ and layer $\ell$, respectively.  Large values of $\Psi_{c, \ell}$ correspond to a community $c$ being large, fast-moving, and/or steeply sloped.

Fig.~\ref{fig:communityvTime} shows a graphical summary of the changes taking place in the overall community structure. Each horiztonal bar represents a multilayer community during the range of dates (layers) it was detected, with the width proportional to $\Psi_{c,\ell}$. We observe 3 persistently strong communities, one each within Mud Creek, Region 1, and Region 2.  Importantly, Mud Creek (outlined in the white polygon) is consistently detected as a strong community over a year before its failure, up through the final data collection. Within Region 2, the community we detect (outlined in the black polygon) corresponds to a creeping landslide that was not detected using InSAR alone; its detection here is an example of how community detection techniques can supplement more traditional ones. Finally, there is a community detected within Region 1 which intermittently appears as an area of concern, but has not been identified as a creeping landslide. Interestingly, no communities were identified within Region 3 because: while it shares similar topographic characteristics to Mud Creek, it is not creeping. Finally, we find that while this type of plot identifies areas of concern, it does not yet provide a framework for forecasting time of failure. In the next section, we develop such a metric. 

\subsection{Forecasting failure via community persistence \label{sec:persistence}} 

We observed that Mud Creek is identified as the same community for the last $\sim$ 400 days in Fig.~\ref{fig:communityvTime}. Mud Creek is one of the few communities that clearly persists through time, while other communities split into smaller or singleton communities. Using these observations, we propose a new measure: \textit{community persistence} $\Pi_\ell$ inspired by \citet{rieck_clique_2018, gonzalez-avella_emergence_2014}. This measures the stability of the nodal composition for each community in relation to the community's size and is defined as
\begin{equation}
    \Pi_\ell = \frac{1}{N}\sum_{c} \frac{c_{\ell-1} \cap c_{\ell}}{n_{c,\ell}},
    \label{eqn:persistence}
\end{equation}
where $N$ is the total number of nodes in the network, $n_{c,\ell}$ is the number of nodes in community $c$ at layer $\ell$, and $c_{\ell-1} \cap c_{\ell}$ is the number of nodes present in community $c$ in both layers $\ell$ and $\ell-1$.  An increase in community persistence indicates that a patch of nodes are consistently being identified as a community $c$. 
This occurs when $c$ is stronger-than-average for consecutive layers; if $c$ were to grow or shrink, then its contribution to $\Pi_\ell$ is smaller than one. The more the communities change membership, the closer $\Pi_{\ell}$ is to zero.

\begin{figure}
    \centering
    \includegraphics[width = \linewidth]{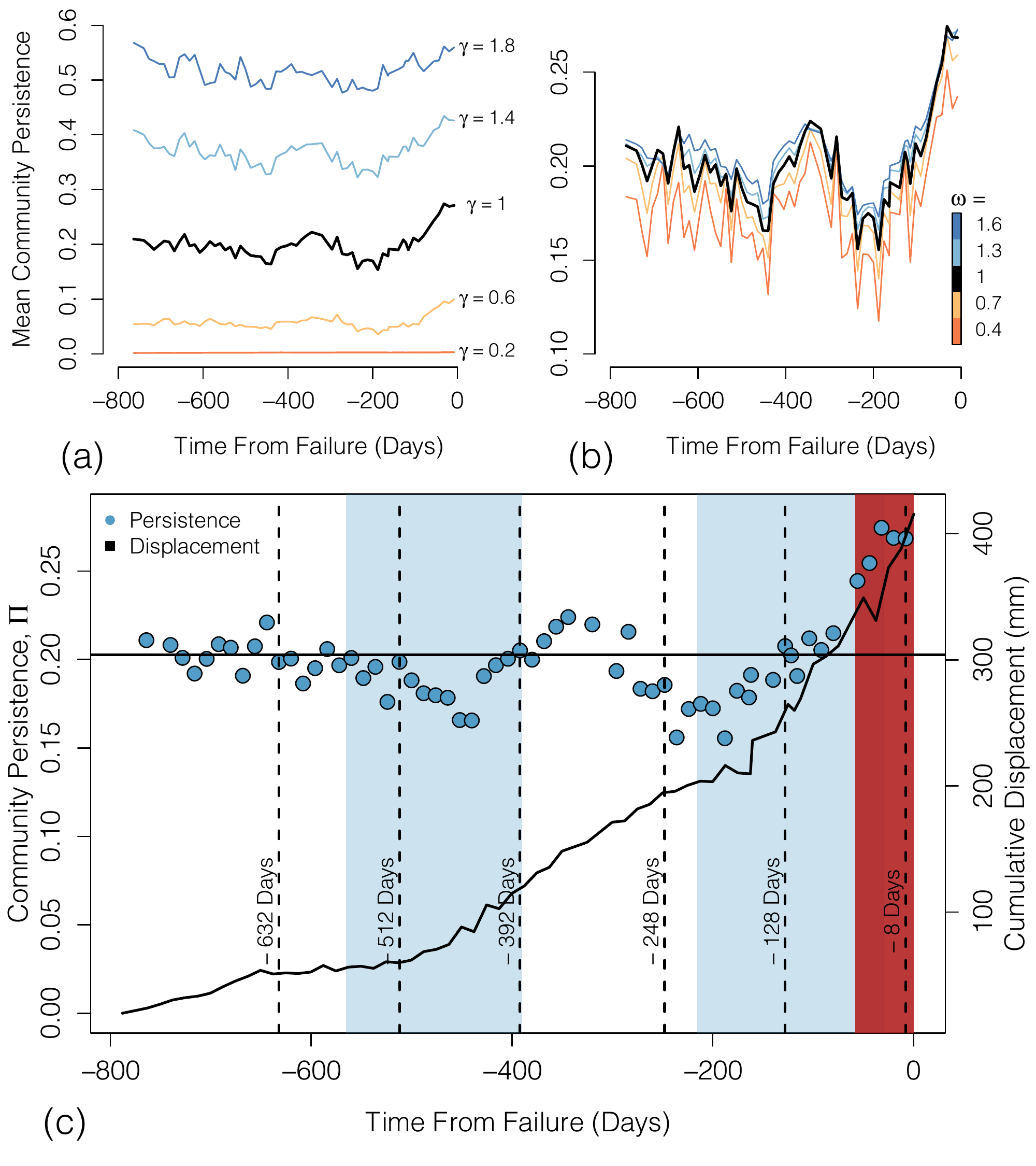}
   \caption{Community persistence $\Pi_\ell$ (Eq.~\ref{eqn:persistence})  as a function of time before failure, averaged over 10 meshes, (a)  for different choices of spatial coupling parameter  $\gamma$ at fixed $\omega = 1$ and (b) for different choices of temporal coupling parameter $\omega$ at fixed  $\gamma = 1.0$) (c) The mean community persistence $\Pi_\ell$ over 10 meshes using $\gamma = 1, \omega = 1$. Standard error, measured across the ensemble of 10 meshes, is smaller than the symbol size; the horizontal black line shows the temporal mean. The dashed lines represent the 6 snapshots of the community structures that are shown in Fig.~\ref{fig:communities}. The blue shaded area represents the wet seasons shown in Fig.~\ref{fig:prcpvel}. The red shaded area highlights where $\Pi_\ell$ starts to significantly rise above the mean. The black line shows the maximum cumulative displacement within the Mud Creek polygon.}
\label{fig:communitypersistence}
\end{figure}

In Fig.~\ref{fig:communitypersistence}(a-b), we show $\Pi_\ell$ for varying values of $\omega$ from 0 to 2 at $\gamma = 1$ and varying values of $\gamma$ from 0 to 2 at $\omega = 1$. Only near $\gamma = 1$ do we observe a signal using $\Pi_\ell$, which confirms the common usage of this value \citep{papadopoulos_evolution_2016, bassett_dynamic_2011}, and corresponds to communities that are stronger-than-average (\textit{i.e.} indicative of a creeping hillslope). This happens because only a few communities are being picked up consistently with a similar subset of nodes from beginning to end. In Fig.~\ref{fig:communitypersistence}(b), we obtain a similar result for values of $\omega \in [0,2]$, but with more noise at low values of $\omega$. Our choice of $\omega = 1$ captures the rising persistence without being hampered by temporal noise (discussed in \S\ref{sec:parameters}). To ensure that our analysis is not sensitive to the ensemble of 10 meshes that we use, we run community detection on 10 randomized meshes, and the mean for those 10 meshes is shown in Fig.~\ref{fig:communitypersistence}.

We see a sharp increase in community persistence (shown in Fig.~\ref{fig:communitypersistence}(c)) about 56 days before failure for all 10 meshes. This is indicative of a group of nodes, each one linked to spatial coordinates, that is continually being established as a community between each layer, in this case the nodes contained within the Mud Creek geometry. As this community becomes stronger (correlated with increasing velocity) and more persistent, we observe that the other communities start to break down into smaller communities, including the community representing the creeping landslide found within Region 2. This happens because the mean edge weight to partition nodes into communities (see Eq.~\ref{eqn:modularity_multi}) is increasing as we get closer to failure (reference Fig.~\ref{fig:velregions}). 

\section{Discussion/Conclusion \label{sec:conclusion}}

We have developed an algorithm that takes a geospatial surface and represents it as a mesh of nodes. A few things we considered when creating the mesh are sensitivity to the mesh we use, artificial results, and good coverage of the area, all of which are addressed by using Poisson disc sampling. To test the spatial resolution of the DEM that is used, we compare a DEM with resolution of 12 m and the DEM we used for network analysis, as described in \S\ref{sec:DEM}, of a higher resolution of 1 m, and note that there are no significant differences between the two, as we downsample the 1 m DEM to 20 m.

We are able to model the dynamics and susceptibility of the system to identify Mud Creek both spatially (through community detection) and temporally (through the persistence of those communities). We do this by incorporating information about both the current state of rheology using velocity data derived from InSAR and the geometry of the area using slope. We found that using only velocity or slope was insufficient  to distinguish creeping landslides from the rest of the area: slope-weighted networks identified watershed basins and velocity-weighted networks identified large communities without separating them into geophysically-relevant areas. Through the inclusion of velocity and slope, we are able to pick up clusters of areas that are strongly-connected to each other. As shown in Fig.~\ref{fig:communityvTime}, we observe three areas of concern: one is Mud Creek, one is a creeping landslide that is already known, and one is a new area of interest for future monitoring. Combined with high-quality InSAR data, network science techniques has allowed us to forecast the location and geometry of the impending failure from 400 days out, without having to visually inspect a map (see \citet{handwerger_shift_2019}).

We quantify these results in Fig.~\ref{fig:communitypersistence}; as we approach time of failure, there is a sharp increase in community persistence $\Pi_\ell$. When compared to the maximum cumulative displacement of the landslide (shown in Fig.~\ref{fig:communitypersistence}), community persistence has a much clearer signal that time of failure is close, although it is not predictive enough to pinpoint a range of days. To test the success of this method using only the first wet season (during which Mud Creek did not fail), we repeated the full analysis using only the first 30 layers (up to $-400$ days). We correctly detected Mud Creek as an area of concern and correctly observed no sharp increase in $\Pi_\ell$, which is consistent with the reality that failure was not imminent. We also note in Fig.~\ref{fig:communitypersistence} that 
the oscillations correspond to the wet/dry seasons, and that the method is not sensitive to choice of a specific mesh from the ensemble. 

This raises several promising avenues for improvement in utility. From the standpoint of continually monitoring, it would be beneficial to instead use pixel tracking displacement or automatically processed InSAR data, instead of InSAR data that has been manually corrected, such as the work of \citet{handwerger_shift_2019}. Previous examination of the InSAR time series by \citet{handwerger_shift_2019} reveal that Mud Creek exhibited strong seasonal displacement driven by precipitation. The displacement increased over the study period due to increasing rainfall during the study period. Although the landslide behavior changed from year to year, and the landslide was moving notably faster in 2017 than previous years, it was concluded by \citet{handwerger_shift_2019} that there was no displacement pattern that is indicative of a slope approaching runaway instability. However, we note here that this is likely a result of bias in the InSAR data itself. InSAR cannot detect large displacements ($>$ half the radar wavelength), which means that if the slope is moving too fast between acquisitions there can be phase bias and an underestimate of true displacement. \citet{warrick_characterizing_2019} showed that the landslide moved over 10 m between March and May 2017, which is too large of displacement for InSAR to measure. From this, we know that the InSAR-measured displacement over the final months prior to collapse is much too slow. Therefore, the quantities used are a lower bound, and improved data would provide for an even stronger signal than is reported here. Significant outlay of human resources would be required to avoid underestimating ground displacements and being dominated by noise, and it remains to be understood what level of spatial, temporal, and velocity resolution is required to achieve the type of results we observed here.

The inclusion of time-delayed rainfall layers, or even approximations to soil moisture (proxy for pore-water pressure) would likely improve the fidelity of the multilayer network in describing the evolving state of the study site. Furthermore, the topography of Earth's surface is constantly changing due to ground deformation, so using a static DEM may introduce challenges and loss of crucial information over a long time period.

While community detection improves the signals of forecasting failure in the Mud Creek area and aids in the identification of creeping landslides, an important open question is how does this method fare on other areas, as well as how to improve the temporal signal of failure.

\section{Acknowledgement}

This work was supported by NSF grant PREEVENTS-566587 and our collaborators on that project. Part of this research was carried out at the Jet Propulsion Laboratory, California Institute of Technology, under a contract with the National Aeronautics and Space Administration (80NM0018D0004).


%apsrev4-2.bst 2019-01-14 (MD) hand-edited version of apsrev4-1.bst
%Control: key (0)
%Control: author (72) initials jnrlst
%Control: editor formatted (1) identically to author
%Control: production of article title (-1) disabled
%Control: page (0) single
%Control: year (1) truncated
%Control: production of eprint (0) enabled
\begin{thebibliography}{0}%
\makeatletter
\providecommand \@ifxundefined [1]{%
 \@ifx{#1\undefined}
}%
\providecommand \@ifnum [1]{%
 \ifnum #1\expandafter \@firstoftwo
 \else \expandafter \@secondoftwo
 \fi
}%
\providecommand \@ifx [1]{%
 \ifx #1\expandafter \@firstoftwo
 \else \expandafter \@secondoftwo
 \fi
}%
\providecommand \natexlab [1]{#1}%
\providecommand \enquote  [1]{``#1''}%
\providecommand \bibnamefont  [1]{#1}%
\providecommand \bibfnamefont [1]{#1}%
\providecommand \citenamefont [1]{#1}%
\providecommand \href@noop [0]{\@secondoftwo}%
\providecommand \href [0]{\begingroup \@sanitize@url \@href}%
\providecommand \@href[1]{\@@startlink{#1}\@@href}%
\providecommand \@@href[1]{\endgroup#1\@@endlink}%
\providecommand \@sanitize@url [0]{\catcode `\\12\catcode `\$12\catcode
  `\&12\catcode `\#12\catcode `\^12\catcode `\_12\catcode `\%12\relax}%
\providecommand \@@startlink[1]{}%
\providecommand \@@endlink[0]{}%
\providecommand \url  [0]{\begingroup\@sanitize@url \@url }%
\providecommand \@url [1]{\endgroup\@href {#1}{\urlprefix }}%
\providecommand \urlprefix  [0]{URL }%
\providecommand \Eprint [0]{\href }%
\providecommand \doibase [0]{https://doi.org/}%
\providecommand \selectlanguage [0]{\@gobble}%
\providecommand \bibinfo  [0]{\@secondoftwo}%
\providecommand \bibfield  [0]{\@secondoftwo}%
\providecommand \translation [1]{[#1]}%
\providecommand \BibitemOpen [0]{}%
\providecommand \bibitemStop [0]{}%
\providecommand \bibitemNoStop [0]{.\EOS\space}%
\providecommand \EOS [0]{\spacefactor3000\relax}%
\providecommand \BibitemShut  [1]{\csname bibitem#1\endcsname}%
\let\auto@bib@innerbib\@empty
%</preamble>
\end{thebibliography}%


%apsrev4-2.bst 2019-01-14 (MD) hand-edited version of apsrev4-1.bst
%Control: key (0)
%Control: author (72) initials jnrlst
%Control: editor formatted (1) identically to author
%Control: production of article title (-1) disabled
%Control: page (0) single
%Control: year (1) truncated
%Control: production of eprint (0) enabled
%


\begin{thebibliography}{57}%
\makeatletter
\providecommand \@ifxundefined [1]{%
 \@ifx{#1\undefined}
}%
\providecommand \@ifnum [1]{%
 \ifnum #1\expandafter \@firstoftwo
 \else \expandafter \@secondoftwo
 \fi
}%
\providecommand \@ifx [1]{%
 \ifx #1\expandafter \@firstoftwo
 \else \expandafter \@secondoftwo
 \fi
}%
\providecommand \natexlab [1]{#1}%
\providecommand \enquote  [1]{``#1''}%
\providecommand \bibnamefont  [1]{#1}%
\providecommand \bibfnamefont [1]{#1}%
\providecommand \citenamefont [1]{#1}%
\providecommand \href@noop [0]{\@secondoftwo}%
\providecommand \href [0]{\begingroup \@sanitize@url \@href}%
\providecommand \@href[1]{\@@startlink{#1}\@@href}%
\providecommand \@@href[1]{\endgroup#1\@@endlink}%
\providecommand \@sanitize@url [0]{\catcode `\\12\catcode `\$12\catcode
  `\&12\catcode `\#12\catcode `\^12\catcode `\_12\catcode `\%12\relax}%
\providecommand \@@startlink[1]{}%
\providecommand \@@endlink[0]{}%
\providecommand \url  [0]{\begingroup\@sanitize@url \@url }%
\providecommand \@url [1]{\endgroup\@href {#1}{\urlprefix }}%
\providecommand \urlprefix  [0]{URL }%
\providecommand \Eprint [0]{\href }%
\providecommand \doibase [0]{https://doi.org/}%
\providecommand \selectlanguage [0]{\@gobble}%
\providecommand \bibinfo  [0]{\@secondoftwo}%
\providecommand \bibfield  [0]{\@secondoftwo}%
\providecommand \translation [1]{[#1]}%
\providecommand \BibitemOpen [0]{}%
\providecommand \bibitemStop [0]{}%
\providecommand \bibitemNoStop [0]{.\EOS\space}%
\providecommand \EOS [0]{\spacefactor3000\relax}%
\providecommand \BibitemShut  [1]{\csname bibitem#1\endcsname}%
\let\auto@bib@innerbib\@empty
%</preamble>
\bibitem [{\citenamefont {Lacroix}\ \emph {et~al.}(2020)\citenamefont
  {Lacroix}, \citenamefont {Handwerger},\ and\ \citenamefont
  {Bièvre}}]{lacroix_life_2020}%
  \BibitemOpen
  \bibfield  {author} {\bibinfo {author} {\bibfnamefont {P.}~\bibnamefont
  {Lacroix}}, \bibinfo {author} {\bibfnamefont {A.~L.}\ \bibnamefont
  {Handwerger}},\ and\ \bibinfo {author} {\bibfnamefont {G.}~\bibnamefont
  {Bièvre}},\ }\href {https://doi.org/10.1038/s43017-020-0072-8} {\bibfield
  {journal} {\bibinfo  {journal} {Nature Reviews Earth \& Environment}\
  }\textbf {\bibinfo {volume} {1}},\ \bibinfo {pages} {404} (\bibinfo {year}
  {2020})}\BibitemShut {NoStop}%
\bibitem [{\citenamefont {Hu}\ \emph {et~al.}(2020)\citenamefont {Hu},
  \citenamefont {Bürgmann}, \citenamefont {Schulz},\ and\ \citenamefont
  {Fielding}}]{hu_four-dimensional_2020}%
  \BibitemOpen
  \bibfield  {author} {\bibinfo {author} {\bibfnamefont {X.}~\bibnamefont
  {Hu}}, \bibinfo {author} {\bibfnamefont {R.}~\bibnamefont {Bürgmann}},
  \bibinfo {author} {\bibfnamefont {W.~H.}\ \bibnamefont {Schulz}},\ and\
  \bibinfo {author} {\bibfnamefont {E.~J.}\ \bibnamefont {Fielding}},\ }\href
  {https://doi.org/10.1038/s41467-020-16617-7} {\bibfield  {journal} {\bibinfo
  {journal} {Nature Communications}\ }\textbf {\bibinfo {volume} {11}},\
  \bibinfo {pages} {2792} (\bibinfo {year} {2020})}\BibitemShut {NoStop}%
\bibitem [{\citenamefont {Rutter}\ and\ \citenamefont
  {Green}(2011)}]{rutter_quantifying_2011}%
  \BibitemOpen
  \bibfield  {author} {\bibinfo {author} {\bibfnamefont {E.}~\bibnamefont
  {Rutter}}\ and\ \bibinfo {author} {\bibfnamefont {S.}~\bibnamefont {Green}},\
  }\href {https://doi.org/10.1144/0016-76492010-133} {\bibfield  {journal}
  {\bibinfo  {journal} {Journal of the Geological Society}\ }\textbf {\bibinfo
  {volume} {168}},\ \bibinfo {pages} {359} (\bibinfo {year}
  {2011})}\BibitemShut {NoStop}%
\bibitem [{\citenamefont {Aryal}\ \emph {et~al.}(2012)\citenamefont {Aryal},
  \citenamefont {Brooks}, \citenamefont {Reid}, \citenamefont {Bawden},\ and\
  \citenamefont {Pawlak}}]{aryal_displacement_2012}%
  \BibitemOpen
  \bibfield  {author} {\bibinfo {author} {\bibfnamefont {A.}~\bibnamefont
  {Aryal}}, \bibinfo {author} {\bibfnamefont {B.~A.}\ \bibnamefont {Brooks}},
  \bibinfo {author} {\bibfnamefont {M.~E.}\ \bibnamefont {Reid}}, \bibinfo
  {author} {\bibfnamefont {G.~W.}\ \bibnamefont {Bawden}},\ and\ \bibinfo
  {author} {\bibfnamefont {G.~R.}\ \bibnamefont {Pawlak}},\ }\bibfield
  {journal} {\bibinfo  {journal} {Journal of Geophysical Research: Earth
  Surface}\ }\textbf {\bibinfo {volume} {117}},\ \href
  {https://doi.org/10.1029/2011JF002161} {10.1029/2011JF002161} (\bibinfo
  {year} {2012})\BibitemShut {NoStop}%
\bibitem [{\citenamefont {Tordesillas}\ \emph {et~al.}(2021)\citenamefont
  {Tordesillas}, \citenamefont {Kahagalage}, \citenamefont {Campbell},
  \citenamefont {Bellett}, \citenamefont {Intrieri},\ and\ \citenamefont
  {Batterham}}]{tordesillas_spatiotemporal_2021}%
  \BibitemOpen
  \bibfield  {author} {\bibinfo {author} {\bibfnamefont {A.}~\bibnamefont
  {Tordesillas}}, \bibinfo {author} {\bibfnamefont {S.}~\bibnamefont
  {Kahagalage}}, \bibinfo {author} {\bibfnamefont {L.}~\bibnamefont
  {Campbell}}, \bibinfo {author} {\bibfnamefont {P.}~\bibnamefont {Bellett}},
  \bibinfo {author} {\bibfnamefont {E.}~\bibnamefont {Intrieri}},\ and\
  \bibinfo {author} {\bibfnamefont {R.}~\bibnamefont {Batterham}},\ }\href
  {https://doi.org/10.1038/s41598-021-88836-x} {\bibfield  {journal} {\bibinfo
  {journal} {Scientific Reports}\ }\textbf {\bibinfo {volume} {11}},\ \bibinfo
  {pages} {9729} (\bibinfo {year} {2021})}\BibitemShut {NoStop}%
\bibitem [{\citenamefont {Carlà}\ \emph {et~al.}(2019)\citenamefont {Carlà},
  \citenamefont {Intrieri}, \citenamefont {Raspini}, \citenamefont {Bardi},
  \citenamefont {Farina}, \citenamefont {Ferretti}, \citenamefont {Colombo},
  \citenamefont {Novali},\ and\ \citenamefont
  {Casagli}}]{carla_perspectives_2019}%
  \BibitemOpen
  \bibfield  {author} {\bibinfo {author} {\bibfnamefont {T.}~\bibnamefont
  {Carlà}}, \bibinfo {author} {\bibfnamefont {E.}~\bibnamefont {Intrieri}},
  \bibinfo {author} {\bibfnamefont {F.}~\bibnamefont {Raspini}}, \bibinfo
  {author} {\bibfnamefont {F.}~\bibnamefont {Bardi}}, \bibinfo {author}
  {\bibfnamefont {P.}~\bibnamefont {Farina}}, \bibinfo {author} {\bibfnamefont
  {A.}~\bibnamefont {Ferretti}}, \bibinfo {author} {\bibfnamefont
  {D.}~\bibnamefont {Colombo}}, \bibinfo {author} {\bibfnamefont
  {F.}~\bibnamefont {Novali}},\ and\ \bibinfo {author} {\bibfnamefont
  {N.}~\bibnamefont {Casagli}},\ }\href
  {https://doi.org/10.1038/s41598-019-50792-y} {\bibfield  {journal} {\bibinfo
  {journal} {Scientific Reports}\ }\textbf {\bibinfo {volume} {9}},\ \bibinfo
  {pages} {1} (\bibinfo {year} {2019})}\BibitemShut {NoStop}%
\bibitem [{\citenamefont {Intrieri}\ \emph {et~al.}(2019)\citenamefont
  {Intrieri}, \citenamefont {Carlà},\ and\ \citenamefont
  {Gigli}}]{intrieri_forecasting_2019}%
  \BibitemOpen
  \bibfield  {author} {\bibinfo {author} {\bibfnamefont {E.}~\bibnamefont
  {Intrieri}}, \bibinfo {author} {\bibfnamefont {T.}~\bibnamefont {Carlà}},\
  and\ \bibinfo {author} {\bibfnamefont {G.}~\bibnamefont {Gigli}},\ }\href
  {https://doi.org/10.1016/j.earscirev.2019.03.019} {\bibfield  {journal}
  {\bibinfo  {journal} {Earth-Science Reviews}\ }\textbf {\bibinfo {volume}
  {193}},\ \bibinfo {pages} {333} (\bibinfo {year} {2019})}\BibitemShut
  {NoStop}%
\bibitem [{\citenamefont {Singh}\ and\ \citenamefont
  {Tordesillas}(2020)}]{singh_spatiotemporal_2020}%
  \BibitemOpen
  \bibfield  {author} {\bibinfo {author} {\bibfnamefont {K.}~\bibnamefont
  {Singh}}\ and\ \bibinfo {author} {\bibfnamefont {A.}~\bibnamefont
  {Tordesillas}},\ }\href {https://doi.org/10.3390/e22010067} {\bibfield
  {journal} {\bibinfo  {journal} {Entropy}\ }\textbf {\bibinfo {volume} {22}},\
  \bibinfo {pages} {67} (\bibinfo {year} {2020})}\BibitemShut {NoStop}%
\bibitem [{\citenamefont {Tordesillas}\ \emph {et~al.}(2018)\citenamefont
  {Tordesillas}, \citenamefont {Zhou},\ and\ \citenamefont
  {Batterham}}]{tordesillas_data-driven_2018}%
  \BibitemOpen
  \bibfield  {author} {\bibinfo {author} {\bibfnamefont {A.}~\bibnamefont
  {Tordesillas}}, \bibinfo {author} {\bibfnamefont {Z.}~\bibnamefont {Zhou}},\
  and\ \bibinfo {author} {\bibfnamefont {R.}~\bibnamefont {Batterham}},\ }\href
  {https://doi.org/10.1016/j.mechrescom.2018.08.008} {\bibfield  {journal}
  {\bibinfo  {journal} {Mechanics Research Communications}\ }\textbf {\bibinfo
  {volume} {92}},\ \bibinfo {pages} {137} (\bibinfo {year} {2018})}\BibitemShut
  {NoStop}%
\bibitem [{\citenamefont {Cascini}\ \emph {et~al.}(2022)\citenamefont
  {Cascini}, \citenamefont {Scoppettuolo},\ and\ \citenamefont
  {Babilio}}]{cascini_forecasting_2022}%
  \BibitemOpen
  \bibfield  {author} {\bibinfo {author} {\bibfnamefont {L.}~\bibnamefont
  {Cascini}}, \bibinfo {author} {\bibfnamefont {M.~R.}\ \bibnamefont
  {Scoppettuolo}},\ and\ \bibinfo {author} {\bibfnamefont {E.}~\bibnamefont
  {Babilio}},\ }\bibfield  {journal} {\bibinfo  {journal} {Landslides}\ }\href
  {https://doi.org/10.1007/s10346-022-01934-3} {10.1007/s10346-022-01934-3}
  (\bibinfo {year} {2022})\BibitemShut {NoStop}%
\bibitem [{\citenamefont {Gabriel}\ \emph {et~al.}(1989)\citenamefont
  {Gabriel}, \citenamefont {Goldstein},\ and\ \citenamefont
  {Zebker}}]{gabriel_mapping_1989}%
  \BibitemOpen
  \bibfield  {author} {\bibinfo {author} {\bibfnamefont {A.~K.}\ \bibnamefont
  {Gabriel}}, \bibinfo {author} {\bibfnamefont {R.~M.}\ \bibnamefont
  {Goldstein}},\ and\ \bibinfo {author} {\bibfnamefont {H.~A.}\ \bibnamefont
  {Zebker}},\ }\href {https://doi.org/10.1029/JB094iB07p09183} {\bibfield
  {journal} {\bibinfo  {journal} {Journal of Geophysical Research}\ }\textbf
  {\bibinfo {volume} {94}},\ \bibinfo {pages} {9183} (\bibinfo {year}
  {1989})}\BibitemShut {NoStop}%
\bibitem [{\citenamefont {Franceschetti}\ and\ \citenamefont
  {Lanari}(2018)}]{franceschetti_synthetic_2018}%
  \BibitemOpen
  \bibfield  {author} {\bibinfo {author} {\bibfnamefont {G.}~\bibnamefont
  {Franceschetti}}\ and\ \bibinfo {author} {\bibfnamefont {R.}~\bibnamefont
  {Lanari}},\ }\href {https://doi.org/10.1201/9780203737484} {\emph {\bibinfo
  {title} {Synthetic {Aperture} {Radar} {Processing}}}},\ \bibinfo {edition}
  {1st}\ ed.\ (\bibinfo  {publisher} {CRC Press},\ \bibinfo {year}
  {2018})\BibitemShut {NoStop}%
\bibitem [{\citenamefont {Di~Traglia}\ \emph {et~al.}(2021)\citenamefont
  {Di~Traglia}, \citenamefont {De~Luca}, \citenamefont {Manzo}, \citenamefont
  {Nolesini}, \citenamefont {Casagli}, \citenamefont {Lanari},\ and\
  \citenamefont {Casu}}]{di_traglia_joint_2021}%
  \BibitemOpen
  \bibfield  {author} {\bibinfo {author} {\bibfnamefont {F.}~\bibnamefont
  {Di~Traglia}}, \bibinfo {author} {\bibfnamefont {C.}~\bibnamefont {De~Luca}},
  \bibinfo {author} {\bibfnamefont {M.}~\bibnamefont {Manzo}}, \bibinfo
  {author} {\bibfnamefont {T.}~\bibnamefont {Nolesini}}, \bibinfo {author}
  {\bibfnamefont {N.}~\bibnamefont {Casagli}}, \bibinfo {author} {\bibfnamefont
  {R.}~\bibnamefont {Lanari}},\ and\ \bibinfo {author} {\bibfnamefont
  {F.}~\bibnamefont {Casu}},\ }\href
  {https://doi.org/10.1016/j.rse.2021.112441} {\bibfield  {journal} {\bibinfo
  {journal} {Remote Sensing of Environment}\ }\textbf {\bibinfo {volume}
  {260}},\ \bibinfo {pages} {112441} (\bibinfo {year} {2021})}\BibitemShut
  {NoStop}%
\bibitem [{\citenamefont {Teshebaeva}\ \emph {et~al.}(2019)\citenamefont
  {Teshebaeva}, \citenamefont {Echtler}, \citenamefont {Bookhagen},\ and\
  \citenamefont {Strecker}}]{teshebaeva_deepseated_2019}%
  \BibitemOpen
  \bibfield  {author} {\bibinfo {author} {\bibfnamefont {K.}~\bibnamefont
  {Teshebaeva}}, \bibinfo {author} {\bibfnamefont {H.~P.}\ \bibnamefont
  {Echtler}}, \bibinfo {author} {\bibfnamefont {B.}~\bibnamefont {Bookhagen}},\
  and\ \bibinfo {author} {\bibfnamefont {M.~R.}\ \bibnamefont {Strecker}},\
  }\href {https://doi.org/10.1002/esp.4648} {\bibfield  {journal} {\bibinfo
  {journal} {Earth Surface Processes and Landforms}\ ,\ \bibinfo {pages}
  {esp.4648}} (\bibinfo {year} {2019})}\BibitemShut {NoStop}%
\bibitem [{\citenamefont {Teshebaeva}\ \emph {et~al.}(2015)\citenamefont
  {Teshebaeva}, \citenamefont {Roessner}, \citenamefont {Echtler},
  \citenamefont {Motagh}, \citenamefont {Wetzel},\ and\ \citenamefont
  {Molodbekov}}]{teshebaeva_alos/palsar_2015}%
  \BibitemOpen
  \bibfield  {author} {\bibinfo {author} {\bibfnamefont {K.}~\bibnamefont
  {Teshebaeva}}, \bibinfo {author} {\bibfnamefont {S.}~\bibnamefont
  {Roessner}}, \bibinfo {author} {\bibfnamefont {H.}~\bibnamefont {Echtler}},
  \bibinfo {author} {\bibfnamefont {M.}~\bibnamefont {Motagh}}, \bibinfo
  {author} {\bibfnamefont {H.-U.}\ \bibnamefont {Wetzel}},\ and\ \bibinfo
  {author} {\bibfnamefont {B.}~\bibnamefont {Molodbekov}},\ }\href
  {https://doi.org/10.3390/rs70708973} {\bibfield  {journal} {\bibinfo
  {journal} {Remote Sensing}\ }\textbf {\bibinfo {volume} {7}},\ \bibinfo
  {pages} {8973} (\bibinfo {year} {2015})}\BibitemShut {NoStop}%
\bibitem [{\citenamefont {Carlà}\ \emph {et~al.}(2018)\citenamefont {Carlà},
  \citenamefont {Farina}, \citenamefont {Intrieri}, \citenamefont {Ketizmen},\
  and\ \citenamefont {Casagli}}]{carla_integration_2018}%
  \BibitemOpen
  \bibfield  {author} {\bibinfo {author} {\bibfnamefont {T.}~\bibnamefont
  {Carlà}}, \bibinfo {author} {\bibfnamefont {P.}~\bibnamefont {Farina}},
  \bibinfo {author} {\bibfnamefont {E.}~\bibnamefont {Intrieri}}, \bibinfo
  {author} {\bibfnamefont {H.}~\bibnamefont {Ketizmen}},\ and\ \bibinfo
  {author} {\bibfnamefont {N.}~\bibnamefont {Casagli}},\ }\href
  {https://doi.org/10.1016/j.enggeo.2018.01.021} {\bibfield  {journal}
  {\bibinfo  {journal} {Engineering Geology}\ }\textbf {\bibinfo {volume}
  {235}},\ \bibinfo {pages} {39} (\bibinfo {year} {2018})}\BibitemShut
  {NoStop}%
\bibitem [{\citenamefont {Bernardi}\ \emph {et~al.}(2021)\citenamefont
  {Bernardi}, \citenamefont {Africa}, \citenamefont {de~Falco}, \citenamefont
  {Formaggia}, \citenamefont {Menafoglio},\ and\ \citenamefont
  {Vantini}}]{bernardi_use_2021}%
  \BibitemOpen
  \bibfield  {author} {\bibinfo {author} {\bibfnamefont {M.~S.}\ \bibnamefont
  {Bernardi}}, \bibinfo {author} {\bibfnamefont {P.~C.}\ \bibnamefont
  {Africa}}, \bibinfo {author} {\bibfnamefont {C.}~\bibnamefont {de~Falco}},
  \bibinfo {author} {\bibfnamefont {L.}~\bibnamefont {Formaggia}}, \bibinfo
  {author} {\bibfnamefont {A.}~\bibnamefont {Menafoglio}},\ and\ \bibinfo
  {author} {\bibfnamefont {S.}~\bibnamefont {Vantini}},\ }\href
  {https://doi.org/10.1007/s11004-021-09948-8} {\bibfield  {journal} {\bibinfo
  {journal} {Mathematical Geosciences}\ }\textbf {\bibinfo {volume} {53}},\
  \bibinfo {pages} {1781} (\bibinfo {year} {2021})}\BibitemShut {NoStop}%
\bibitem [{\citenamefont {Dick}\ \emph {et~al.}(2015)\citenamefont {Dick},
  \citenamefont {Eberhardt}, \citenamefont {Cabrejo-Liévano}, \citenamefont
  {Stead},\ and\ \citenamefont {Rose}}]{dick_development_2015}%
  \BibitemOpen
  \bibfield  {author} {\bibinfo {author} {\bibfnamefont {G.~J.}\ \bibnamefont
  {Dick}}, \bibinfo {author} {\bibfnamefont {E.}~\bibnamefont {Eberhardt}},
  \bibinfo {author} {\bibfnamefont {A.~G.}\ \bibnamefont {Cabrejo-Liévano}},
  \bibinfo {author} {\bibfnamefont {D.}~\bibnamefont {Stead}},\ and\ \bibinfo
  {author} {\bibfnamefont {N.~D.}\ \bibnamefont {Rose}},\ }\href
  {https://doi.org/10.1139/cgj-2014-0028} {\bibfield  {journal} {\bibinfo
  {journal} {Canadian Geotechnical Journal}\ }\textbf {\bibinfo {volume}
  {52}},\ \bibinfo {pages} {515} (\bibinfo {year} {2015})}\BibitemShut
  {NoStop}%
\bibitem [{\citenamefont {Zhou}\ \emph {et~al.}(2022)\citenamefont {Zhou},
  \citenamefont {Tordesillas}, \citenamefont {Intrieri}, \citenamefont
  {Di~Traglia}, \citenamefont {Qian},\ and\ \citenamefont
  {Catani}}]{zhou_pinpointing_2022}%
  \BibitemOpen
  \bibfield  {author} {\bibinfo {author} {\bibfnamefont {S.}~\bibnamefont
  {Zhou}}, \bibinfo {author} {\bibfnamefont {A.}~\bibnamefont {Tordesillas}},
  \bibinfo {author} {\bibfnamefont {E.}~\bibnamefont {Intrieri}}, \bibinfo
  {author} {\bibfnamefont {F.}~\bibnamefont {Di~Traglia}}, \bibinfo {author}
  {\bibfnamefont {G.}~\bibnamefont {Qian}},\ and\ \bibinfo {author}
  {\bibfnamefont {F.}~\bibnamefont {Catani}},\ }\bibfield  {journal} {\bibinfo
  {journal} {Journal of Geophysical Research: Solid Earth}\ }\textbf {\bibinfo
  {volume} {127}},\ \href {https://doi.org/10.1029/2021JB022957}
  {10.1029/2021JB022957} (\bibinfo {year} {2022})\BibitemShut {NoStop}%
\bibitem [{\citenamefont {Das}\ and\ \citenamefont
  {Tordesillas}(2019)}]{das_near_2019}%
  \BibitemOpen
  \bibfield  {author} {\bibinfo {author} {\bibfnamefont {S.}~\bibnamefont
  {Das}}\ and\ \bibinfo {author} {\bibfnamefont {A.}~\bibnamefont
  {Tordesillas}},\ }\href {https://doi.org/10.3390/rs11232777} {\bibfield
  {journal} {\bibinfo  {journal} {Remote Sensing}\ }\textbf {\bibinfo {volume}
  {11}},\ \bibinfo {pages} {2777} (\bibinfo {year} {2019})}\BibitemShut
  {NoStop}%
\bibitem [{\citenamefont {Agliardi}\ \emph {et~al.}(2020)\citenamefont
  {Agliardi}, \citenamefont {Scuderi}, \citenamefont {Fusi},\ and\
  \citenamefont {Collettini}}]{agliardi_slow--fast_2020}%
  \BibitemOpen
  \bibfield  {author} {\bibinfo {author} {\bibfnamefont {F.}~\bibnamefont
  {Agliardi}}, \bibinfo {author} {\bibfnamefont {M.~M.}\ \bibnamefont
  {Scuderi}}, \bibinfo {author} {\bibfnamefont {N.}~\bibnamefont {Fusi}},\ and\
  \bibinfo {author} {\bibfnamefont {C.}~\bibnamefont {Collettini}},\ }\href
  {https://doi.org/10.1038/s41467-020-15093-3} {\bibfield  {journal} {\bibinfo
  {journal} {Nature Communications}\ }\textbf {\bibinfo {volume} {11}},\
  \bibinfo {pages} {1352} (\bibinfo {year} {2020})}\BibitemShut {NoStop}%
\bibitem [{\citenamefont {Torres}\ \emph {et~al.}(2021)\citenamefont {Torres},
  \citenamefont {Blevins}, \citenamefont {Bassett},\ and\ \citenamefont
  {Eliassi-Rad}}]{torres_why_2021}%
  \BibitemOpen
  \bibfield  {author} {\bibinfo {author} {\bibfnamefont {L.}~\bibnamefont
  {Torres}}, \bibinfo {author} {\bibfnamefont {A.~S.}\ \bibnamefont {Blevins}},
  \bibinfo {author} {\bibfnamefont {D.}~\bibnamefont {Bassett}},\ and\ \bibinfo
  {author} {\bibfnamefont {T.}~\bibnamefont {Eliassi-Rad}},\ }\href
  {https://doi.org/10.1137/20M1355896} {\bibfield  {journal} {\bibinfo
  {journal} {SIAM Review}\ }\textbf {\bibinfo {volume} {63}},\ \bibinfo {pages}
  {435} (\bibinfo {year} {2021})}\BibitemShut {NoStop}%
\bibitem [{\citenamefont {Le~Bouil}\ \emph {et~al.}(2014)\citenamefont
  {Le~Bouil}, \citenamefont {Amon}, \citenamefont {McNamara},\ and\
  \citenamefont {Crassous}}]{le_bouil_emergence_2014}%
  \BibitemOpen
  \bibfield  {author} {\bibinfo {author} {\bibfnamefont {A.}~\bibnamefont
  {Le~Bouil}}, \bibinfo {author} {\bibfnamefont {A.}~\bibnamefont {Amon}},
  \bibinfo {author} {\bibfnamefont {S.}~\bibnamefont {McNamara}},\ and\
  \bibinfo {author} {\bibfnamefont {J.}~\bibnamefont {Crassous}},\ }\href
  {https://doi.org/10.1103/PhysRevLett.112.246001} {\bibfield  {journal}
  {\bibinfo  {journal} {Physical Review Letters}\ }\textbf {\bibinfo {volume}
  {112}},\ \bibinfo {pages} {246001} (\bibinfo {year} {2014})}\BibitemShut
  {NoStop}%
\bibitem [{\citenamefont {Fazelpour}(2022)}]{fazelpour_failure_2022}%
  \BibitemOpen
  \bibfield  {author} {\bibinfo {author} {\bibfnamefont {F.}~\bibnamefont
  {Fazelpour}},\ }\emph {\bibinfo {title} {Failure and {Flow} in {Sheared}
  {Granular} {Materials}}},\ \href
  {https://www.lib.ncsu.edu/resolver/1840.20/39806} {\bibinfo {type} {{PhD}
  {Thesis}}},\ \bibinfo  {school} {North Carolina State University} (\bibinfo
  {year} {2022})\BibitemShut {NoStop}%
\bibitem [{\citenamefont {Mei}\ \emph {et~al.}(2022)\citenamefont {Mei},
  \citenamefont {Ma}, \citenamefont {Wang}, \citenamefont {Wu},\ and\
  \citenamefont {Zhou}}]{mei_micro-_2022}%
  \BibitemOpen
  \bibfield  {author} {\bibinfo {author} {\bibfnamefont {J.}~\bibnamefont
  {Mei}}, \bibinfo {author} {\bibfnamefont {G.}~\bibnamefont {Ma}}, \bibinfo
  {author} {\bibfnamefont {Q.}~\bibnamefont {Wang}}, \bibinfo {author}
  {\bibfnamefont {T.}~\bibnamefont {Wu}},\ and\ \bibinfo {author}
  {\bibfnamefont {W.}~\bibnamefont {Zhou}},\ }\href
  {https://doi.org/10.1016/j.ijsolstr.2022.111763} {\bibfield  {journal}
  {\bibinfo  {journal} {International Journal of Solids and Structures}\
  }\textbf {\bibinfo {volume} {251}},\ \bibinfo {pages} {111763} (\bibinfo
  {year} {2022})}\BibitemShut {NoStop}%
\bibitem [{\citenamefont {Tordesillas}(2007)}]{tordesillas_force_2007}%
  \BibitemOpen
  \bibfield  {author} {\bibinfo {author} {\bibfnamefont {A.}~\bibnamefont
  {Tordesillas}},\ }\href {https://doi.org/10.1080/14786430701594848}
  {\bibfield  {journal} {\bibinfo  {journal} {Philosophical Magazine}\ }\textbf
  {\bibinfo {volume} {87}},\ \bibinfo {pages} {4987} (\bibinfo {year}
  {2007})}\BibitemShut {NoStop}%
\bibitem [{\citenamefont {Handwerger}\ \emph {et~al.}(2019)\citenamefont
  {Handwerger}, \citenamefont {Huang}, \citenamefont {Fielding}, \citenamefont
  {Booth},\ and\ \citenamefont {Bürgmann}}]{handwerger_shift_2019}%
  \BibitemOpen
  \bibfield  {author} {\bibinfo {author} {\bibfnamefont {A.~L.}\ \bibnamefont
  {Handwerger}}, \bibinfo {author} {\bibfnamefont {M.-H.}\ \bibnamefont
  {Huang}}, \bibinfo {author} {\bibfnamefont {E.~J.}\ \bibnamefont {Fielding}},
  \bibinfo {author} {\bibfnamefont {A.~M.}\ \bibnamefont {Booth}},\ and\
  \bibinfo {author} {\bibfnamefont {R.}~\bibnamefont {Bürgmann}},\ }\href
  {https://doi.org/10.1038/s41598-018-38300-0} {\bibfield  {journal} {\bibinfo
  {journal} {Scientific Reports}\ }\textbf {\bibinfo {volume} {9}},\ \bibinfo
  {pages} {1569} (\bibinfo {year} {2019})}\BibitemShut {NoStop}%
\bibitem [{\citenamefont {Roering}\ \emph {et~al.}(2015)\citenamefont
  {Roering}, \citenamefont {Mackey}, \citenamefont {Handwerger}, \citenamefont
  {Booth}, \citenamefont {Schmidt}, \citenamefont {Bennett},\ and\
  \citenamefont {Cerovski-Darriau}}]{roering_beyond_2015}%
  \BibitemOpen
  \bibfield  {author} {\bibinfo {author} {\bibfnamefont {J.~J.}\ \bibnamefont
  {Roering}}, \bibinfo {author} {\bibfnamefont {B.~H.}\ \bibnamefont {Mackey}},
  \bibinfo {author} {\bibfnamefont {A.~L.}\ \bibnamefont {Handwerger}},
  \bibinfo {author} {\bibfnamefont {A.~M.}\ \bibnamefont {Booth}}, \bibinfo
  {author} {\bibfnamefont {D.~A.}\ \bibnamefont {Schmidt}}, \bibinfo {author}
  {\bibfnamefont {G.~L.}\ \bibnamefont {Bennett}},\ and\ \bibinfo {author}
  {\bibfnamefont {C.}~\bibnamefont {Cerovski-Darriau}},\ }\href
  {https://doi.org/10.1016/j.geomorph.2015.02.013} {\bibfield  {journal}
  {\bibinfo  {journal} {Geomorphology}\ }\textbf {\bibinfo {volume} {236}},\
  \bibinfo {pages} {109} (\bibinfo {year} {2015})}\BibitemShut {NoStop}%
\bibitem [{\citenamefont {Handwerger}\ \emph {et~al.}(2015)\citenamefont
  {Handwerger}, \citenamefont {Roering}, \citenamefont {Schmidt},\ and\
  \citenamefont {Rempel}}]{handwerger_kinematics_2015}%
  \BibitemOpen
  \bibfield  {author} {\bibinfo {author} {\bibfnamefont {A.~L.}\ \bibnamefont
  {Handwerger}}, \bibinfo {author} {\bibfnamefont {J.~J.}\ \bibnamefont
  {Roering}}, \bibinfo {author} {\bibfnamefont {D.~A.}\ \bibnamefont
  {Schmidt}},\ and\ \bibinfo {author} {\bibfnamefont {A.~W.}\ \bibnamefont
  {Rempel}},\ }\href {https://doi.org/10.1016/j.geomorph.2015.06.003}
  {\bibfield  {journal} {\bibinfo  {journal} {Geomorphology}\ }\textbf
  {\bibinfo {volume} {246}},\ \bibinfo {pages} {321} (\bibinfo {year}
  {2015})}\BibitemShut {NoStop}%
\bibitem [{\citenamefont {Scheingross}\ \emph {et~al.}(2013)\citenamefont
  {Scheingross}, \citenamefont {Minchew}, \citenamefont {Mackey}, \citenamefont
  {Simons}, \citenamefont {Lamb},\ and\ \citenamefont
  {Hensley}}]{scheingross_fault-zone_2013}%
  \BibitemOpen
  \bibfield  {author} {\bibinfo {author} {\bibfnamefont {J.~S.}\ \bibnamefont
  {Scheingross}}, \bibinfo {author} {\bibfnamefont {B.~M.}\ \bibnamefont
  {Minchew}}, \bibinfo {author} {\bibfnamefont {B.~H.}\ \bibnamefont {Mackey}},
  \bibinfo {author} {\bibfnamefont {M.}~\bibnamefont {Simons}}, \bibinfo
  {author} {\bibfnamefont {M.~P.}\ \bibnamefont {Lamb}},\ and\ \bibinfo
  {author} {\bibfnamefont {S.}~\bibnamefont {Hensley}},\ }\href
  {https://doi.org/10.1130/B30719.1} {\bibfield  {journal} {\bibinfo  {journal}
  {Geological Society of America Bulletin}\ }\textbf {\bibinfo {volume}
  {125}},\ \bibinfo {pages} {473} (\bibinfo {year} {2013})}\BibitemShut
  {NoStop}%
\bibitem [{\citenamefont {Warrick}\ \emph {et~al.}(2019)\citenamefont
  {Warrick}, \citenamefont {Ritchie}, \citenamefont {Schmidt}, \citenamefont
  {Reid},\ and\ \citenamefont {Logan}}]{warrick_characterizing_2019}%
  \BibitemOpen
  \bibfield  {author} {\bibinfo {author} {\bibfnamefont {J.~A.}\ \bibnamefont
  {Warrick}}, \bibinfo {author} {\bibfnamefont {A.~C.}\ \bibnamefont
  {Ritchie}}, \bibinfo {author} {\bibfnamefont {K.~M.}\ \bibnamefont
  {Schmidt}}, \bibinfo {author} {\bibfnamefont {M.~E.}\ \bibnamefont {Reid}},\
  and\ \bibinfo {author} {\bibfnamefont {J.}~\bibnamefont {Logan}},\ }\href
  {https://doi.org/10.1007/s10346-019-01160-4} {\bibfield  {journal} {\bibinfo
  {journal} {Landslides}\ }\textbf {\bibinfo {volume} {16}},\ \bibinfo {pages}
  {1201} (\bibinfo {year} {2019})}\BibitemShut {NoStop}%
\bibitem [{noa(2015)}]{noauthor_prism_2015}%
  \BibitemOpen
  {\bibinfo {title} {{PRISM} {Climate}
  {Group}, {Oregon} {State} {U}}} (\bibinfo {year} {2015}),\ \bibinfo {note}
  {accessed on 2022-01-13} \url {https://prism.oregonstate.edu/} \BibitemShut {NoStop}%
\bibitem [{\citenamefont {Swain}\ \emph {et~al.}(2018)\citenamefont {Swain},
  \citenamefont {Langenbrunner}, \citenamefont {Neelin},\ and\ \citenamefont
  {Hall}}]{swain_increasing_2018}%
  \BibitemOpen
  \bibfield  {author} {\bibinfo {author} {\bibfnamefont {D.~L.}\ \bibnamefont
  {Swain}}, \bibinfo {author} {\bibfnamefont {B.}~\bibnamefont
  {Langenbrunner}}, \bibinfo {author} {\bibfnamefont {J.~D.}\ \bibnamefont
  {Neelin}},\ and\ \bibinfo {author} {\bibfnamefont {A.}~\bibnamefont {Hall}},\
  }\href {https://doi.org/10.1038/s41558-018-0140-y} {\bibfield  {journal}
  {\bibinfo  {journal} {Nature Climate Change}\ }\textbf {\bibinfo {volume}
  {8}},\ \bibinfo {pages} {427} (\bibinfo {year} {2018})}\BibitemShut {NoStop}%
\bibitem [{\citenamefont {Allen}\ and\ \citenamefont
  {Luptowitz}(2017)}]{allen_nino-like_2017}%
  \BibitemOpen
  \bibfield  {author} {\bibinfo {author} {\bibfnamefont {R.~J.}\ \bibnamefont
  {Allen}}\ and\ \bibinfo {author} {\bibfnamefont {R.}~\bibnamefont
  {Luptowitz}},\ }\href {https://doi.org/10.1038/ncomms16055} {\bibfield
  {journal} {\bibinfo  {journal} {Nature Communications}\ }\textbf {\bibinfo
  {volume} {8}},\ \bibinfo {pages} {16055} (\bibinfo {year}
  {2017})}\BibitemShut {NoStop}%
\bibitem [{\citenamefont {Iverson}\ \emph {et~al.}(2000)\citenamefont
  {Iverson}, \citenamefont {Reid}, \citenamefont {Iverson}, \citenamefont
  {LaHusen}, \citenamefont {Logan}, \citenamefont {Mann},\ and\ \citenamefont
  {Brien}}]{iverson_acute_2000}%
  \BibitemOpen
  \bibfield  {author} {\bibinfo {author} {\bibfnamefont {R.~M.}\ \bibnamefont
  {Iverson}}, \bibinfo {author} {\bibfnamefont {M.~E.}\ \bibnamefont {Reid}},
  \bibinfo {author} {\bibfnamefont {N.~R.}\ \bibnamefont {Iverson}}, \bibinfo
  {author} {\bibfnamefont {R.~G.}\ \bibnamefont {LaHusen}}, \bibinfo {author}
  {\bibfnamefont {M.}~\bibnamefont {Logan}}, \bibinfo {author} {\bibfnamefont
  {J.~E.}\ \bibnamefont {Mann}},\ and\ \bibinfo {author} {\bibfnamefont
  {D.~L.}\ \bibnamefont {Brien}},\ }\href
  {https://doi.org/10.1126/science.290.5491.513} {\bibfield  {journal}
  {\bibinfo  {journal} {Science}\ }\textbf {\bibinfo {volume} {290}},\ \bibinfo
  {pages} {513} (\bibinfo {year} {2000})}\BibitemShut {NoStop}%
\bibitem [{\citenamefont {Mitarai}\ and\ \citenamefont
  {Nori}(2006)}]{mitarai_wet_2006}%
  \BibitemOpen
  \bibfield  {author} {\bibinfo {author} {\bibfnamefont {N.}~\bibnamefont
  {Mitarai}}\ and\ \bibinfo {author} {\bibfnamefont {F.}~\bibnamefont {Nori}},\
  }\href {https://doi.org/10.1080/00018730600626065} {\bibfield  {journal}
  {\bibinfo  {journal} {Advances in Physics}\ }\textbf {\bibinfo {volume}
  {55}},\ \bibinfo {pages} {1} (\bibinfo {year} {2006})}\BibitemShut {NoStop}%
\bibitem [{\citenamefont {Schulz}\ \emph {et~al.}(2009)\citenamefont {Schulz},
  \citenamefont {McKenna}, \citenamefont {Kibler},\ and\ \citenamefont
  {Biavati}}]{schulz_relations_2009}%
  \BibitemOpen
  \bibfield  {author} {\bibinfo {author} {\bibfnamefont {W.~H.}\ \bibnamefont
  {Schulz}}, \bibinfo {author} {\bibfnamefont {J.~P.}\ \bibnamefont {McKenna}},
  \bibinfo {author} {\bibfnamefont {J.~D.}\ \bibnamefont {Kibler}},\ and\
  \bibinfo {author} {\bibfnamefont {G.}~\bibnamefont {Biavati}},\ }\href
  {https://doi.org/10.1007/s10346-009-0157-4} {\bibfield  {journal} {\bibinfo
  {journal} {Landslides}\ }\textbf {\bibinfo {volume} {6}},\ \bibinfo {pages}
  {181} (\bibinfo {year} {2009})}\BibitemShut {NoStop}%
\bibitem [{\citenamefont {Lu}\ and\ \citenamefont
  {Godt}(2013)}]{lu_hillslope_2013}%
  \BibitemOpen
  \bibfield  {author} {\bibinfo {author} {\bibfnamefont {N.}~\bibnamefont
  {Lu}}\ and\ \bibinfo {author} {\bibfnamefont {J.}~\bibnamefont {Godt}},\
  }\href {https://doi.org/10.1017/CBO9781139108164} {\emph {\bibinfo {title}
  {Hillslope {Hydrology} and {Stability}}}}\ (\bibinfo  {publisher} {Cambridge
  University Press},\ \bibinfo {address} {Cambridge},\ \bibinfo {year}
  {2013})\BibitemShut {NoStop}%
\bibitem [{\citenamefont {Girard}\ \emph {et~al.}(2010)\citenamefont {Girard},
  \citenamefont {Amitrano},\ and\ \citenamefont {Weiss}}]{girard_failure_2010}%
  \BibitemOpen
  \bibfield  {author} {\bibinfo {author} {\bibfnamefont {L.}~\bibnamefont
  {Girard}}, \bibinfo {author} {\bibfnamefont {D.}~\bibnamefont {Amitrano}},\
  and\ \bibinfo {author} {\bibfnamefont {J.}~\bibnamefont {Weiss}},\ }\href
  {https://doi.org/10.1088/1742-5468/2010/01/P01013} {\bibfield  {journal}
  {\bibinfo  {journal} {Journal of Statistical Mechanics: Theory and
  Experiment}\ }\textbf {\bibinfo {volume} {2010}},\ \bibinfo {pages} {P01013}
  (\bibinfo {year} {2010})}\BibitemShut {NoStop}%
\bibitem [{\citenamefont {Jaeger}\ \emph {et~al.}(2007)\citenamefont {Jaeger},
  \citenamefont {Cook},\ and\ \citenamefont
  {Zimmerman}}]{jaeger_fundamentals_2007}%
  \BibitemOpen
  \bibfield  {author} {\bibinfo {author} {\bibfnamefont {J.~C.}\ \bibnamefont
  {Jaeger}}, \bibinfo {author} {\bibfnamefont {N.~G.~W.}\ \bibnamefont
  {Cook}},\ and\ \bibinfo {author} {\bibfnamefont {R.~W.}\ \bibnamefont
  {Zimmerman}},\ }\href@noop {} {\emph {\bibinfo {title} {Fundamentals of rock
  mechanics}}},\ \bibinfo {edition} {4th}\ ed.\ (\bibinfo  {publisher}
  {Blackwell Pub},\ \bibinfo {address} {Malden, MA},\ \bibinfo {year}
  {2007})\BibitemShut {NoStop}%
\bibitem [{\citenamefont {Palla}\ \emph {et~al.}(2007)\citenamefont {Palla},
  \citenamefont {Barabási},\ and\ \citenamefont
  {Vicsek}}]{palla_quantifying_2007}%
  \BibitemOpen
  \bibfield  {author} {\bibinfo {author} {\bibfnamefont {G.}~\bibnamefont
  {Palla}}, \bibinfo {author} {\bibfnamefont {A.-L.}\ \bibnamefont
  {Barabási}},\ and\ \bibinfo {author} {\bibfnamefont {T.}~\bibnamefont
  {Vicsek}},\ }\href {https://doi.org/10.1038/nature05670} {\bibfield
  {journal} {\bibinfo  {journal} {Nature}\ }\textbf {\bibinfo {volume} {446}},\
  \bibinfo {pages} {664} (\bibinfo {year} {2007})}\BibitemShut {NoStop}%
\bibitem [{\citenamefont {Whitaker}\ \emph {et~al.}(2019)\citenamefont
  {Whitaker}, \citenamefont {Varga}, \citenamefont {Hsiao}, \citenamefont
  {Solomon}, \citenamefont {Swan},\ and\ \citenamefont
  {Furst}}]{whitaker_colloidal_2019}%
  \BibitemOpen
  \bibfield  {author} {\bibinfo {author} {\bibfnamefont {K.~A.}\ \bibnamefont
  {Whitaker}}, \bibinfo {author} {\bibfnamefont {Z.}~\bibnamefont {Varga}},
  \bibinfo {author} {\bibfnamefont {L.~C.}\ \bibnamefont {Hsiao}}, \bibinfo
  {author} {\bibfnamefont {M.~J.}\ \bibnamefont {Solomon}}, \bibinfo {author}
  {\bibfnamefont {J.~W.}\ \bibnamefont {Swan}},\ and\ \bibinfo {author}
  {\bibfnamefont {E.~M.}\ \bibnamefont {Furst}},\ }\href
  {https://doi.org/10.1038/s41467-019-10039-w} {\bibfield  {journal} {\bibinfo
  {journal} {Nature Communications}\ }\textbf {\bibinfo {volume} {10}},\
  \bibinfo {pages} {2237} (\bibinfo {year} {2019})}\BibitemShut {NoStop}%
\bibitem [{\citenamefont {Fortunato}(2010)}]{fortunato_community_2010}%
  \BibitemOpen
  \bibfield  {author} {\bibinfo {author} {\bibfnamefont {S.}~\bibnamefont
  {Fortunato}},\ }\href {https://doi.org/10.1016/j.physrep.2009.11.002}
  {\bibfield  {journal} {\bibinfo  {journal} {Physics Reports}\ }\textbf
  {\bibinfo {volume} {486}},\ \bibinfo {pages} {75} (\bibinfo {year} {2010})},\
  \bibinfo {note} {arXiv: 0906.0612}\BibitemShut {NoStop}%
\bibitem [{\citenamefont {Papadopoulos}\ \emph {et~al.}(2016)\citenamefont
  {Papadopoulos}, \citenamefont {Puckett}, \citenamefont {Daniels},\ and\
  \citenamefont {Bassett}}]{papadopoulos_evolution_2016}%
  \BibitemOpen
  \bibfield  {author} {\bibinfo {author} {\bibfnamefont {L.}~\bibnamefont
  {Papadopoulos}}, \bibinfo {author} {\bibfnamefont {J.~G.}\ \bibnamefont
  {Puckett}}, \bibinfo {author} {\bibfnamefont {K.~E.}\ \bibnamefont
  {Daniels}},\ and\ \bibinfo {author} {\bibfnamefont {D.~S.}\ \bibnamefont
  {Bassett}},\ }\href {https://doi.org/10.1103/PhysRevE.94.032908} {\bibfield
  {journal} {\bibinfo  {journal} {Physical Review E}\ }\textbf {\bibinfo
  {volume} {94}},\ \bibinfo {pages} {032908} (\bibinfo {year}
  {2016})}\BibitemShut {NoStop}%
\bibitem [{noa(2010)}]{noauthor_usgs_2010}%
  \BibitemOpen
  \bibfield  {author} {\bibinfo {author} {\bibfnamefont {U.~S.}~\bibnamefont
  {Geological~Survey}}\ (\bibinfo {year}
  {2014})},\
  {\bibinfo {title} {ARRA-CA CentralCoast-Z4 2010,}\ } \bibinfo {note}\url {https://portal.opentopography.org/usgsDataset?dsid=ARRA-CA_CentralCoast-Z4_2010},\ {last accessed on 2021-05-17}\BibitemShut {NoStop}%
\bibitem [{\citenamefont {Cook}(1986)}]{cook_stochastic_1986}%
  \BibitemOpen
  \bibfield  {author} {\bibinfo {author} {\bibfnamefont {R.~L.}\ \bibnamefont
  {Cook}},\ }\href {https://doi.org/10.1145/7529.8927} {\bibfield  {journal}
  {\bibinfo  {journal} {ACM Transactions on Graphics}\ }\textbf {\bibinfo
  {volume} {5}},\ \bibinfo {pages} {51} (\bibinfo {year} {1986})}\BibitemShut
  {NoStop}%
\bibitem [{\citenamefont {Porter}\ \emph {et~al.}(2009)\citenamefont {Porter},
  \citenamefont {Onnela},\ and\ \citenamefont
  {Mucha}}]{porter_communities_2009}%
  \BibitemOpen
  \bibfield  {author} {\bibinfo {author} {\bibfnamefont {M.~A.}\ \bibnamefont
  {Porter}}, \bibinfo {author} {\bibfnamefont {J.-P.}\ \bibnamefont {Onnela}},\
  and\ \bibinfo {author} {\bibfnamefont {P.~J.}\ \bibnamefont {Mucha}},\
  }\href@noop {} {\bibfield  {journal} {\bibinfo  {journal} {Notices of the
  American Mathematical Society}\ }\textbf {\bibinfo {volume} {56}},\ \bibinfo
  {pages} {1082} (\bibinfo {year} {2009})}\BibitemShut {NoStop}%
\bibitem [{\citenamefont {Mucha}\ \emph {et~al.}(2010)\citenamefont {Mucha},
  \citenamefont {Richardson}, \citenamefont {Macon}, \citenamefont {Porter},\
  and\ \citenamefont {Onnela}}]{mucha_community_2010}%
  \BibitemOpen
  \bibfield  {author} {\bibinfo {author} {\bibfnamefont {P.~J.}\ \bibnamefont
  {Mucha}}, \bibinfo {author} {\bibfnamefont {T.}~\bibnamefont {Richardson}},
  \bibinfo {author} {\bibfnamefont {K.}~\bibnamefont {Macon}}, \bibinfo
  {author} {\bibfnamefont {M.~A.}\ \bibnamefont {Porter}},\ and\ \bibinfo
  {author} {\bibfnamefont {J.-P.}\ \bibnamefont {Onnela}},\ }\href
  {http://www.jstor.org/stable/40655926} {\bibfield  {journal} {\bibinfo
  {journal} {Science}\ }\textbf {\bibinfo {volume} {328}},\ \bibinfo {pages}
  {876} (\bibinfo {year} {2010})}\BibitemShut {NoStop}%
\bibitem [{\citenamefont {Newman}\ and\ \citenamefont
  {Girvan}(2004)}]{newman_finding_2004}%
  \BibitemOpen
  \bibfield  {author} {\bibinfo {author} {\bibfnamefont {M.~E.~J.}\
  \bibnamefont {Newman}}\ and\ \bibinfo {author} {\bibfnamefont
  {M.}~\bibnamefont {Girvan}},\ }\href
  {https://doi.org/10.1103/PhysRevE.69.026113} {\bibfield  {journal} {\bibinfo
  {journal} {Physical Review E}\ }\textbf {\bibinfo {volume} {69}},\ \bibinfo
  {pages} {026113} (\bibinfo {year} {2004})}\BibitemShut {NoStop}%
\bibitem [{\citenamefont {Newman}(2006)}]{newman_modularity_2006}%
  \BibitemOpen
  \bibfield  {author} {\bibinfo {author} {\bibfnamefont {M.~E.~J.}\
  \bibnamefont {Newman}},\ }\href {https://doi.org/10.1073/pnas.0601602103}
  {\bibfield  {journal} {\bibinfo  {journal} {Proceedings of the National
  Academy of Sciences}\ }\textbf {\bibinfo {volume} {103}},\ \bibinfo {pages}
  {8577} (\bibinfo {year} {2006})}\BibitemShut {NoStop}%
\bibitem [{\citenamefont {Jeub}\ \emph {et~al.}(1 19)\citenamefont {Jeub},
  \citenamefont {Bazzi}, \citenamefont {Jutla},\ and\ \citenamefont
  {Mucha}}]{netwiki}%
  \BibitemOpen
  \bibfield  {author} {\bibinfo {author} {\bibfnamefont {L.~G.~S.}\
  \bibnamefont {Jeub}}, \bibinfo {author} {\bibfnamefont {M.}~\bibnamefont
  {Bazzi}}, \bibinfo {author} {\bibfnamefont {I.~S.}\ \bibnamefont {Jutla}},\
  and\ \bibinfo {author} {\bibfnamefont {P.~J.}\ \bibnamefont {Mucha}},\ }{\bibinfo {title} {A generalized
  louvain method for community detection implemented in matlab}} (\bibinfo
  {year} {2011-19}),\ \bibinfo {note} {downloaded on 2020-05-15} \url{http://netwiki.amath.unc.edu/GenLouvain, https://github.com/GenLouvain/GenLouvain}\BibitemShut
  {NoStop}%
\bibitem [{\citenamefont {Sarzynska}\ \emph {et~al.}(2016)\citenamefont
  {Sarzynska}, \citenamefont {Leicht}, \citenamefont {Chowell},\ and\
  \citenamefont {Porter}}]{sarzynska_null_2016}%
  \BibitemOpen
  \bibfield  {author} {\bibinfo {author} {\bibfnamefont {M.}~\bibnamefont
  {Sarzynska}}, \bibinfo {author} {\bibfnamefont {E.~A.}\ \bibnamefont
  {Leicht}}, \bibinfo {author} {\bibfnamefont {G.}~\bibnamefont {Chowell}},\
  and\ \bibinfo {author} {\bibfnamefont {M.~A.}\ \bibnamefont {Porter}},\
  }\href {https://doi.org/10.1093/comnet/cnv027} {\bibfield  {journal}
  {\bibinfo  {journal} {Journal of Complex Networks}\ }\textbf {\bibinfo
  {volume} {4}},\ \bibinfo {pages} {363} (\bibinfo {year} {2016})}\BibitemShut
  {NoStop}%
\bibitem [{\citenamefont {Bassett}\ \emph {et~al.}(2015)\citenamefont
  {Bassett}, \citenamefont {Owens}, \citenamefont {Porter}, \citenamefont
  {Manning},\ and\ \citenamefont {Daniels}}]{bassett_extraction_2015}%
  \BibitemOpen
  \bibfield  {author} {\bibinfo {author} {\bibfnamefont {D.~S.}\ \bibnamefont
  {Bassett}}, \bibinfo {author} {\bibfnamefont {E.~T.}\ \bibnamefont {Owens}},
  \bibinfo {author} {\bibfnamefont {M.~A.}\ \bibnamefont {Porter}}, \bibinfo
  {author} {\bibfnamefont {M.~L.}\ \bibnamefont {Manning}},\ and\ \bibinfo
  {author} {\bibfnamefont {K.~E.}\ \bibnamefont {Daniels}},\ }\href
  {https://doi.org/10.1039/C4SM01821D} {\bibfield  {journal} {\bibinfo
  {journal} {Soft Matter}\ }\textbf {\bibinfo {volume} {11}},\ \bibinfo {pages}
  {2731} (\bibinfo {year} {2015})}\BibitemShut {NoStop}%
\bibitem [{\citenamefont {Bassett}\ \emph {et~al.}(2011)\citenamefont
  {Bassett}, \citenamefont {Wymbs}, \citenamefont {Porter}, \citenamefont
  {Mucha}, \citenamefont {Carlson},\ and\ \citenamefont
  {Grafton}}]{bassett_dynamic_2011}%
  \BibitemOpen
  \bibfield  {author} {\bibinfo {author} {\bibfnamefont {D.~S.}\ \bibnamefont
  {Bassett}}, \bibinfo {author} {\bibfnamefont {N.~F.}\ \bibnamefont {Wymbs}},
  \bibinfo {author} {\bibfnamefont {M.~A.}\ \bibnamefont {Porter}}, \bibinfo
  {author} {\bibfnamefont {P.~J.}\ \bibnamefont {Mucha}}, \bibinfo {author}
  {\bibfnamefont {J.~M.}\ \bibnamefont {Carlson}},\ and\ \bibinfo {author}
  {\bibfnamefont {S.~T.}\ \bibnamefont {Grafton}},\ }\href
  {https://doi.org/10.1073/pnas.1018985108} {\bibfield  {journal} {\bibinfo
  {journal} {Proceedings of the National Academy of Sciences}\ }\textbf
  {\bibinfo {volume} {108}},\ \bibinfo {pages} {7641} (\bibinfo {year}
  {2011})}\BibitemShut {NoStop}%
\bibitem [{\citenamefont {Bassett}\ \emph {et~al.}(2013)\citenamefont
  {Bassett}, \citenamefont {Porter}, \citenamefont {Wymbs}, \citenamefont
  {Grafton}, \citenamefont {Carlson},\ and\ \citenamefont
  {Mucha}}]{bassett_robust_2013}%
  \BibitemOpen
  \bibfield  {author} {\bibinfo {author} {\bibfnamefont {D.~S.}\ \bibnamefont
  {Bassett}}, \bibinfo {author} {\bibfnamefont {M.~A.}\ \bibnamefont {Porter}},
  \bibinfo {author} {\bibfnamefont {N.~F.}\ \bibnamefont {Wymbs}}, \bibinfo
  {author} {\bibfnamefont {S.~T.}\ \bibnamefont {Grafton}}, \bibinfo {author}
  {\bibfnamefont {J.~M.}\ \bibnamefont {Carlson}},\ and\ \bibinfo {author}
  {\bibfnamefont {P.~J.}\ \bibnamefont {Mucha}},\ }\href
  {https://doi.org/10.1063/1.4790830} {\bibfield  {journal} {\bibinfo
  {journal} {Chaos: An Interdisciplinary Journal of Nonlinear Science}\
  }\textbf {\bibinfo {volume} {23}},\ \bibinfo {pages} {013142} (\bibinfo
  {year} {2013})}\BibitemShut {NoStop}%
\bibitem [{\citenamefont {Rieck}\ \emph {et~al.}(2018)\citenamefont {Rieck},
  \citenamefont {Fugacci}, \citenamefont {Lukasczyk},\ and\ \citenamefont
  {Leitte}}]{rieck_clique_2018}%
  \BibitemOpen
  \bibfield  {author} {\bibinfo {author} {\bibfnamefont {B.}~\bibnamefont
  {Rieck}}, \bibinfo {author} {\bibfnamefont {U.}~\bibnamefont {Fugacci}},
  \bibinfo {author} {\bibfnamefont {J.}~\bibnamefont {Lukasczyk}},\ and\
  \bibinfo {author} {\bibfnamefont {H.}~\bibnamefont {Leitte}},\ }\href
  {https://doi.org/10.1109/TVCG.2017.2744321} {\bibfield  {journal} {\bibinfo
  {journal} {IEEE Transactions on Visualization and Computer Graphics}\
  }\textbf {\bibinfo {volume} {24}},\ \bibinfo {pages} {822} (\bibinfo {year}
  {2018})}\BibitemShut {NoStop}%
\bibitem [{\citenamefont {González-Avella}\ \emph {et~al.}(2014)\citenamefont
  {González-Avella}, \citenamefont {Cosenza}, \citenamefont {Herrera},\ and\
  \citenamefont {Tucci}}]{gonzalez-avella_emergence_2014}%
  \BibitemOpen
  \bibfield  {author} {\bibinfo {author} {\bibfnamefont {J.~C.}\ \bibnamefont
  {González-Avella}}, \bibinfo {author} {\bibfnamefont {M.~G.}\ \bibnamefont
  {Cosenza}}, \bibinfo {author} {\bibfnamefont {J.~L.}\ \bibnamefont
  {Herrera}},\ and\ \bibinfo {author} {\bibfnamefont {K.}~\bibnamefont
  {Tucci}},\ }\href {https://doi.org/10.1209/0295-5075/107/28002} {\bibfield
  {journal} {\bibinfo  {journal} {EPL (Europhysics Letters)}\ }\textbf
  {\bibinfo {volume} {107}},\ \bibinfo {pages} {28002} (\bibinfo {year}
  {2014})}\BibitemShut {NoStop}%
 \bibitem [{\citenamefont {Desai}()}]{github_desai}%
  \BibitemOpen Numerical methods are published at
  \bibfield  {author} {\bibinfo {author} {\bibfnamefont {V.~D.}\ \bibnamefont
  {Desai}},\ }{\bibinfo {title} {networkLandslide},\ }\url {https://github.com/vddesai-97/networkLandslide}\BibitemShut {NoStop}%
  
\end{thebibliography}
\end{document}